\documentclass[useAMS,a4paper,fleqn,usenatbib]{mnras}
\usepackage{color}
\usepackage{graphicx}	% Including figure files
\usepackage{amsmath}	% Advanced maths commands
\usepackage{amssymb}	% Extra maths symbols
\usepackage{multicol}        % Multi-column entries in tables
\usepackage{bm}		% Bold maths symbols, including upright Greek
\usepackage{pdflscape}	% Landscape pages

\usepackage[T1]{fontenc}
\usepackage{ae,aecompl}

\usepackage{txfonts}

\newcommand{\xmm}{{\em XMM--Newton}}
\newcommand{\nus}{{\em NuSTAR}}
\newcommand{\chandra}{{\em Chandra}}
\newcommand{\suz}{{\em Suzaku}}

\newcommand{\sax}{{\em BeppoSAX}}
\newcommand{\swift}{{\em Swift}}
\newcommand{\vlba}{{\em VLBA}}
\newcommand{\fermi}{{\em Fermi}}

\newcommand{\aer}[3]{$#1^{+ #2}_{- #3}$}
\newcommand{\aerm}[3]{#1^{+ #2}_{- #3}}
\newcommand{\aerexp}[4]{$#1^{+ #2}_{- #3} \times 10^{#4}$}
\newcommand{\ser}[2]{$#1 \pm #2$}
\newcommand{\serm}[2]{#1 \pm #2}
\newcommand{\serexp}[3]{($#1 \pm #2) \times 10^{#3}$}
\newcommand{\lsup}[1]{$< #1$}
\newcommand{\linf}[1]{$> #1$}
\newcommand{\expo}[2]{$ #1 \times 10^{#2}$}

\newcommand{\pers}{s$^{-1}$}
\newcommand{\chisq}{\chi^{2}}
\newcommand{\rchisq}{\chi^{2}/\textrm{dof}}
\newcommand{\dchi}{\Delta \chi^{2}}
\newcommand{\ddof}{\Delta \textrm{dof}}
\newcommand{\dcash}{\Delta C}
\newcommand{\rcash}{C/\textrm{dof}}

\newcommand{\nh}{N_{\textrm{H}}}

\newcommand{\nhone}{N_{\textrm{H,}\textsc{wa1}}}
\newcommand{\nhtwo}{N_{\textrm{H,}\textsc{wa2}}}
\newcommand{\xione}{\xi_{\textsc{wa1}}}
\newcommand{\xitwo}{\xi_{\textsc{wa2}}}

\newcommand{\vone}{v_{\textsc{wa1}}}
\newcommand{\vtwo}{v_{\textsc{wa2}}}
\newcommand{\afe}{A_{\textrm{Fe}}}
\newcommand{\mbh}{M_{\textrm{BH}}}

\newcommand{\ftwoten}{F_{\textrm{2--10keV}}}
\newcommand{\lbol}{L_{\textrm{bol}}}

\newcommand{\fek}{Fe~K$\alpha$}

\newcommand{\xspec}{{\sc xspec}}

\newcommand{\relxill}{{\sc relxill}}

\newcommand{\nthcomp}{{\sc nthcomp}}
\newcommand{\mytorus}{{\sc mytorus}}

\newcommand{\cloudy}{{\sc cloudy}}

\newcommand{\kdblur}{{\sc kdblur}}
\newcommand{\pexrav}{{\sc pexrav}}
\newcommand{\redden}{{\sc redden}}
\newcommand{\smallbb}{{\sc smallBB}}

\newcommand{\kte}{kT_{\textrm{e}}}
\newcommand{\ktbb}{kT_{\textrm{BB}}}
\newcommand{\kteh}{kT_{\textrm{e,h}}}
\newcommand{\ktew}{kT_{\textrm{e,w}}}
\newcommand{\gammah}{\Gamma_{\textrm{h}}}
\newcommand{\gammaw}{\Gamma_{\textrm{w}}}

\newcommand{\fluxcgs}{ergs~s$^{-1}$~cm$^{-2}$}
\newcommand{\lumcgs}{ergs~s$^{-1}$}
\newcommand{\kms}{km~s$^{-1}$}
\newcommand{\vturb}{\sigma_v}
\newcommand{\ovii}{O~{\sc vii}}

\newcommand{\feii}{Fe~{\sc ii}}
\newcommand{\sqcm}{cm$^{-2}$}
\newcommand{\rin}{R_{\textrm{in}}}
\newcommand{\rg}{$R_{\textrm{G}}$}

\newcommand{\tc}{3C~382}
\newcommand{\ngc}{NGC~4593}

\usepackage[space]{grffile}
\usepackage[normalem]{ulem}
\usepackage{hyperref}
\usepackage{comment}
\bibpunct{(}{)}{;}{a}{}{,}

\begin{document}
\title[Radio/X-ray monitoring of 3C 382. High-energy view]{Radio/X-ray monitoring of the broad-line radio galaxy 3C 382. High-energy view with \xmm\ and \nus}

\author[F. Ursini et al.]
	{F. Ursini,$^{1}$\thanks{e-mail: \href{mailto:ursini@iasfbo.inaf.it}{\texttt{ursini@iasfbo.inaf.it}}}
	P.-O. Petrucci,$^{2}$
	G. Matt,$^{3}$
	S. Bianchi,$^{3}$
	M. Cappi,$^{1}$
	M. Dadina,$^{1}$ 
	P. Grandi,$^{1}$
	\newauthor
	E. Torresi,$^{1}$
	D. R. Ballantyne,$^4$
	B. De Marco,$^5$
	A. De Rosa,$^6$
	M. Giroletti,$^7$
	J.~Malzac,$^{8}$
	\newauthor
	A.~Marinucci,$^{3}$
	R.~Middei,$^3$
	G. Ponti,$^{9}$
	and
	A. Tortosa$^{3}$
\\
	$^1$ INAF-Osservatorio di astrofisica e scienza dello spazio di Bologna, Via Piero Gobetti 93/3, 40129 Bologna, Italy. \\
	$^2$ Univ. Grenoble Alpes, IPAG, F-38000 Grenoble, France. \\
	$^3$ Dipartimento di Matematica e Fisica, Universit\`a degli Studi Roma Tre, via della Vasca Navale 84, 00146 Roma, Italy. \\
	$^4$ Center for Relativistic Astrophysics, School of Physics, Georgia Institute of Technology, Atlanta, GA 30332, USA. \\
	$^5$ Nicolaus Copernicus Astronomical Center, PL-00-716 Warsaw, Poland.\\
%%	%\and
	$^6$ INAF/Istituto di Astrofisica e Planetologia Spaziali, via Fosso del Cavaliere, 00133 Roma, Italy.\\
	$^7$ INAF-Istituto di Radioastronomia, Via Gobetti 101, I-40129 Bologna, Italy.\\
%%	%\and 
	$^8$ IRAP, Universit\'{e} de Toulouse, CNRS, UPS, CNES, Toulouse, France.\\
%%	%\and
	$^{9}$ Max-Planck-Institut f\"ur extraterrestrische Physik, Giessenbachstrasse, D-85748 Garching, Germany. \\
}

\date{Released Xxxx Xxxxx XX}

\maketitle

\label{firstpage}

\begin{abstract}
We present the analysis of five joint \xmm/\nus\ observations, 20 ks each and separated by 12 days, of the broad-line radio galaxy \tc. The data were obtained as part of a campaign performed in September-October 2016 simultaneously with \vlba.
The radio data and their relation with the X-ray ones
will be discussed in a following paper.
The source exhibits a moderate flux variability in the UV/X-ray bands, and a limited spectral variability especially in the soft X-ray band.
In agreement with past observations, we find the presence of a warm absorber, an iron K$\alpha$ line with no associated Compton reflection hump, and a variable soft excess well described by a thermal Comptonization component. The data are consistent with a ``two-corona'' scenario, in which the UV emission and soft excess are produced by a warm ($kT \simeq 0.6$ keV), optically thick ($\tau \simeq 20$) corona  consistent with being a slab fully covering a nearly passive accretion disc, while the hard X-ray emission is due to a hot corona intercepting roughly 10\% of the soft emission. 
These results are remarkably similar to those generally found in radio-quiet Seyferts, thus suggesting a common accretion mechanism.
\end{abstract}


\section{Introduction}
Active galactic nuclei (AGNs) are powerful emitters over several decades of frequency. 
Their central engine is believed to be a supermassive black hole surrounded by an accretion disc, which mostly emits in the optical/UV band. The X-ray emission is thought to originate, at least in radio-quiet AGNs, via Comptonization of disc photons in a hot corona located in the inner region \cite[e.g.][]{haardt&maraschi1991,hmg1994,hmg1997}. 
The primary X-ray emission from the hot corona can be modified by the interaction with the surrounding matter. In particular, it can be absorbed by neutral or ionized gas (the so-called warm absorber), and Compton reflected by the disc \cite[e.g.][]{george&fabian1991,MPP1991} or by more distant material such as the molecular torus at pc scales \cite[e.g.][]{ghm1994,matt2003}.
Moreover, an excess of emission on top of the extrapolated high-energy power law is commonly observed in the spectra of AGNs below 1-2 keV \cite[e.g.][]{walter&fink1993,caixa1}. The origin of this so-called ``soft excess'' is still uncertain \cite[e.g.][]{done2012SE}. It could be due to a blend of several relativistically blurred emission lines from an ionized disc \cite[e.g.][]{ross&fabian1993,crummy2006,ponti2006,walton2013}. Alternatively, the soft excess could be the high-energy tail of the Comptonized emission from a ``warm'' plasma \cite[e.g.][]{mag_5548,pop2013mrk509,rozenn2014mrk509SE,matt2014ark120,porquet2018,middei2018}. According to this interpretation, the optical/UV to soft X-ray emission would be produced in a warm ($\kte\sim 1$ keV), optically thick ($\tau \sim 10-20$) corona covering a nearly passive disc, which only reprocesses the X-ray emission from the corona \cite[]{rozanska2015,cheeses}.

AGNs can also be strong particle accelerators, leading to relativistic jets producing radio through gamma radiation \cite[``jetted'' AGNs,][]{padovani2017}.
Following \cite{fanaroff&riley}, radio galaxies are divided into two morphology and radio power subclasses: the
low-luminosity Fanaroff-Riley (FR) I and the high-luminosity FR II. 
Generally FR Is are found to have low accretion rates and/or low radiative efficiency, and their X-ray emission is likely jet-related \cite[e.g.][]{balmaverde2006,hardcastle2009,mingo2014}. On the other hand, the X-ray emission of FR IIs is most likely accretion-related \cite[e.g.][]{grandi2006}.
The diversity in the observational properties of radio-loud AGNs (e.g. radio morphology, optical and X-ray spectra) can be explained by unification schemes, as a result of anisotropy and orientation effects \cite[e.g.][]{barthel1989,urry&padovani,tadhunter2016}.
However, our general understanding of the central engine of AGNs mostly derives from radio-quiet sources. This is in part due to the lower number density of radio-loud AGNs, which are roughly 10-20\% of the total, meaning that there are fewer bright objects \cite[e.g.][]{urry&padovani}. The emission of radio-loud sources can also be rather complex because of the broad-band jet component \cite[e.g.][]{sambruna2004,worrall2005,grandi2006}. As a result, the dichotomy between radio-quiet and radio-loud AGNs is still debated \cite[e.g.][]{sikora2007,orienti2015}. 
A physical connection likely exists between the accretion flow and the jet activity in AGNs, as indicated by the existence of the so-called fundamental plane of black hole activity \cite[]{merloni2003,falcke2004}, but the underlying mechanism is a matter of speculation. For example, the X-ray corona could actually be the base of the jet \cite[e.g.][]{markoff2005}. 
Another possibility could be the so-called jet emitting discs \cite[][]{ferreira2006} which are currently applied to X-ray binaries \cite[e.g.][]{pop2010,marcel2018}.

To investigate the relation between accretion and ejection mechanisms in AGNs, broad-line radio galaxies (BLRGs) are generally considered ideal targets, for two main reasons. First, the jet of BLRGs does not dominate the spectral emission by pointing directly towards the observer (as in blazars). Second, they are generally not obscured in X-rays, i.e. they are analogous to Seyfert 1 galaxies in this respect.
For example, multiwavelength studies on the BLRGs 3C~120 \cite[]{marscher2002,ogle2005,chatt2009,lohfink2013} and 3C~111 \cite[]{chatt2011} have revealed a relationship between events in the radio jet and the X-ray emission, and even the gamma-ray emission in the case of 3C~111 \cite[]{grandi2012}. In particular, the ejection of bright, superluminal knots in the radio jet are preceded by significant dips in the X-ray light curve, indicating a physical connection between the accretion disc and the jet \cite[e.g.][]{chatt2009,chatt2011}. This might also suggest an analogy with black hole X-ray binaries \cite[e.g.][]{marscher2002,fender2006}, 
in particular with the low/hard or intermediate state of these sources (while jets are not observed during the high/soft state).

We performed
a joint monitoring program with the Very Long Baseline Array (\vlba), \xmm\ and \nus\ on \tc, a nearby ($z=0.058$) BLRG hosting a supermassive black hole of \serexp{1.0}{0.3}{9} solar masses \cite[from reverberation mapping,][]{fausnaugh2017}. This is the first monitoring of a BLRG performed by the X-ray satellites \xmm\ and \nus\ jointly with the \vlba. 
In this paper, we focus on the UV to hard X-rays emission, while the radio properties and the connection with the high-energy emission will be discussed in a forthcoming work.

Past X-ray observations of \tc\ with \textit{Ginga}, \textit{ASCA} and \textit{EXOSAT} revealed the presence of a moderate \fek\ emission line and of a soft excess \cite[]{wozniak1998}. From a \sax\ observation in 1998, \cite{grandi2001} found a high-energy cut-off at $\sim 120$ keV, weak reprocessed features and a soft excess not explained by extended thermal emission. 
Analogous results were obtained from a long \textit{RXTE}/\chandra\ observation in 2004, with indications of a softer-when-brighter behaviour similar to that of radio-quiet Seyferts \cite[]{gliozzi2007}. From a \suz\ observation in 2007, \cite{sambruna2011} reported the detection of a relativistically broadened \fek\ line, a bump above 10 keV and a soft excess. All these features were found to be consistent with a common origin from ionized reflection \cite[]{sambruna2011}. From \xmm\ data taken in 2008, \cite{torresi2010} reported the detection of a warm absorber, observed for the first time in a BLRG. \nus\ observed \tc\ in 2012 (simultaneously with \swift) and in 2013, during different flux states \cite[by a factor of 1.7:][]{ballantyne2014}. 
\cite{ballantyne2014} reported different values of the coronal parameters in the two observations, the corona being cooler ($\kte = \aerm{231}{50}{88}$ keV) in the higher flux state and hotter ($\kte = \serm{330}{30}$ keV) in the lower flux state. This behaviour is consistent with that observed in radio-quiet Seyferts \cite[e.g.][]{lubinski2010} and indicates that the main X-ray-emitting region is the corona, rather than the jet \cite[]{ballantyne2014}. Finally, there was no detection of a reflection hump, although the \nus\ spectra showed a \fek\ emission line consistent with originating from a radius larger than 50 gravitational radii \cite[\rg;][]{ballantyne2014,grandi2001}.
Concerning the radio properties, \tc\ has a FR II morphology, and it exhibits a $1.68'$-long jet north-east of the core and two radio lobes, with a total extension of $3'$ \cite[]{black1992}. The total flux density at 8.4 GHz is 190 mJy with a compact core of 115 mJy, from Very Long Baseline Interferometric (\textit{VLBI}) imaging \cite[][]{giova1994}. The jet inclination and velocity are estimated to be $<45$ deg and $>0.6 c$ respectively, from the jet to counterjet brightness ratio \cite[]{giova2001}. From the high-energy properties, however, the jet contribution to the X-ray continuum is likely small \cite[see also][]{grandi2007}. \tc\ is also not detected in gamma-rays with \fermi\ \cite[]{kataoka2011,hooper2016}. 

This paper is organized as follows. In Section \ref{sec:obs}, we describe
the UV/X-ray observations and data reduction. In Section \ref{sec:timing} we discuss the timing properties. In Section \ref{sec:x} we present the analysis of the \xmm\ and \nus\ spectra. In Section \ref{sec:discussion} we discuss the results and summarize our conclusions.

\section{Observations and data reduction}\label{sec:obs}
\tc\ was observed five times by \xmm\ \cite[]{xmm} and \nus\ \cite[]{harrison2013nustar} between 2016 August 29 and 2016 October 17. Each pointing had a net exposure of ${\sim} 20$ ks. 
The log of the data sets is reported in Table \ref{tab:log}. 

\begin{table}
	\begin{center}
		%\scriptsize
		\caption{Logs of the \xmm\ and \nus\ observations of the source. \label{tab:log}}
		\begin{tabular}{ c c c c c} 
			\hline Obs. & Observatories &  Obs. Id. & Start time (\textsc{utc})  & Net exp.\\ 
			& & & yyyy-mm-dd & (ks)  \\ \hline 
			1 
			& \xmm  & 0790600101  & 2016-08-29 & 20 \\ 
			& \nus  
			& 60202015002 &  & 23 \\
			%& \vlba\ & BU034A & & \\
			 \hline 
			2
			& \xmm  & 0790600201  & 2016-09-11 & 15 \\ 
			& \nus  
			& 60202015004 &  & 24 \\
			%& \vlba\ & BU034B & & \\
			\hline 
			3
			& \xmm  & 0790600301  & 2016-09-22 & 19 \\ 
			& \nus  
			& 60202015006 &  & 21 \\
			%& \vlba\ & BU034C & & \\
			\hline 
			4
			& \xmm  & 0790600401  & 2016-10-05 & 15 \\ 
			& \nus  
			& 60202015008 &  & 22 \\
			%& \vlba\ & BU034D & & \\
			\hline 		
			5
			& \xmm  & 0790600501  & 2016-10-17 & 16 \\ 
			& \nus  
			& 60202015010 &  & 21 \\
			%& \vlba\ & BU034E & & \\
			\hline 		
		\end{tabular}
	\end{center}
\end{table} 

\xmm\ observed the source with the optical monitor \cite[OM;][]{OM}, the EPIC cameras \cite[][]{pn,MOS} and the Reflection Grating Spectrometer \cite[RGS;][]{RGS}.
The data were processed using the \xmm\ Science Analysis System (\textsc{sas} v16.1).
The OM photometric filters were operated in the Science User Defined image/fast mode. The images were taken with the U, UVW1, UVM2, and UVW2 filters, with an exposure time of 4.4 ks for each image. The OM data were processed with the \textsc{sas} pipeline \textsc{omichain}, and prepared for the spectral analysis using the \textsc{sas} task \textsc{om2pha}.
The EPIC instruments were operating in the Small Window mode, with the thin filter applied. 
Given the much higher effective area of the pn detector compared with MOS, throughout the paper we discuss results obtained from pn data. 
However, the spectral parameters are consistent among MOS and pn (albeit with larger uncertainties in MOS).
Source extraction radii and screening for high-background intervals were determined through an iterative process that maximizes the signal-to-noise ratio \cite[][]{pico2004}. The background was extracted from circular regions with a radius of 50 arcsec, while the source extraction radii were allowed to be in the range 20--40 arcsec; the best extraction radius was in every case found to be 40 arcsec. The light curves were corrected and background-subtracted using the \textsc{sas} task \textsc{epiclcorr}. The EPIC-pn spectra were grouped such that each spectral bin contained at least 30 counts, and not oversampling the spectral resolution by a factor greater than 3. Finally, the RGS data were extracted using the standard \textsc{sas} task \textsc{rgsproc}. 

The \nus\ data were reduced using the standard pipeline (\textsc{nupipeline}) in the \nus\ Data Analysis Software (\textsc{nustardas}, v1.9.3), using calibration files from \nus\ {\sc caldb} v20171002. Spectra and light curves were extracted using the standard tool {\sc nuproducts} for each of the two hard X-ray detectors aboard \nus, sitting inside the corresponding focal plane modules A and B (FPMA and FPMB). The source data were extracted from circular regions with a radius of 75 arcsec, and background was extracted from a blank area close to the source. The spectra were binned to have a signal-to-noise ratio greater than 5 in each spectral channel, and not oversampling the instrumental resolution by a factor greater than 2.5. The spectra from FPMA and FPMB were analysed jointly, but not combined.

\section{Timing properties}\label{sec:timing}
\begin{figure*} 
	\includegraphics[width=2\columnwidth]{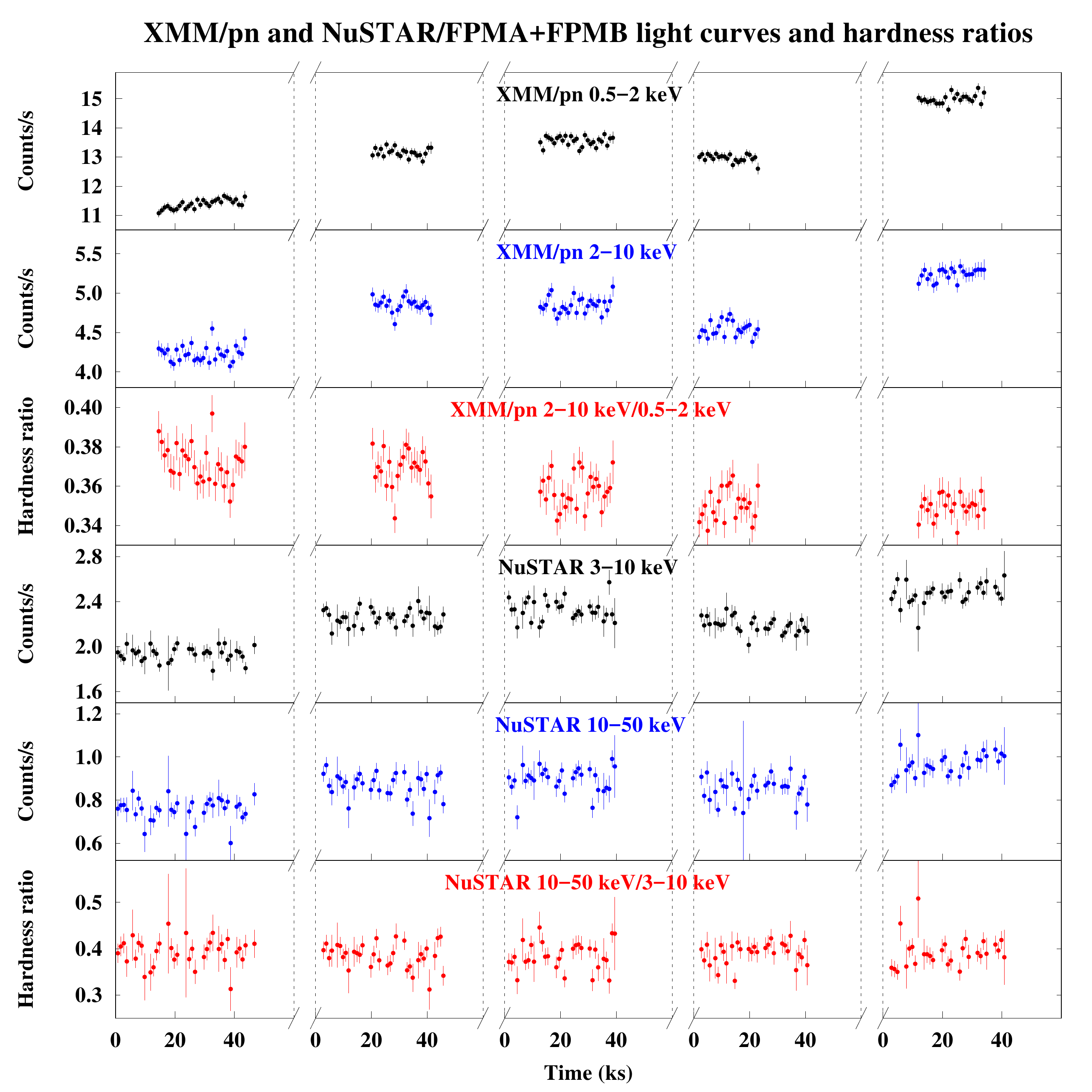}
	\caption{\label{fig:lc} Light curves of the five joint \xmm\ and \nus\ observations of \tc. The exposures are spaced by 11-12 d. Time bins of 1 ks are used. Top panel: \xmm/pn count rate light curve in the 0.5--2 keV band. Second panel: \xmm/pn count rate light curve in the 2--10 keV band. Third panel: \xmm/pn hardness ratio (2--10/0.5--2 keV) light curve. Fourth panel: \nus\ count rate light curve in the 3--10 keV band (FPMA and FPMB data are co-added). Fifth panel: \nus\ count rate light curve in the 10--50 keV band. Bottom panel: \nus\ hardness ratio (10--50/3--10 keV) light curve.}
\end{figure*}

In Fig. \ref{fig:lc} we plot the light curves of \xmm/pn and \nus\ in different energy ranges. The source exhibits a moderate flux variability between different observations, up to ${\sim} 30\%$ in the 0.5--2 keV band. We also plot the pn (2--10 keV)/(0.5--2 keV) hardness ratio and the \nus\ (10--50 keV)/(3--10 keV) hardness ratio. These light curves show a weak spectral variability between different observations, the soft band being the most variable. 

A convenient tool for investigating the flux variability is the normalised excess variance \cite[e.g.][]{nandra1997,vaughan2003,caixa3}, defined as:
\begin{equation}
\sigma^2_{\rm{rms}}=\frac{1}{N\mu^2}\sum_{i=1}^{N} \left[ (X_i-\mu)^2-\sigma_i^2 \right]
\end{equation}
where $N$ is the number of good time bins in a segment of the light curve, $\mu$ is the unweighted mean of the count rate within that segment, $X_i$ represents the count rate and $\sigma_i^2$ is the associated uncertainty. We computed the normalised excess variance in the 2-10 keV band for all the observations of our campaign, using 20 ks time bins, obtaining $\sigma^2_{\rm{rms}} < 4 \times10^{-4}$.
A correlation between $\sigma_{\rm{rms}}^2$ and the black hole mass $\mbh$ is well established in radio-quiet Seyferts \cite[e.g.][]{caixa3}.
Since the bulk of the X-ray continuum of \tc\ most likely originates in a thermal, Seyfert-like corona \cite[]{grandi2001,sambruna2011,ballantyne2014},
we can estimate a lower limit to the black hole mass assuming the $\sigma_{\rm{rms}}^2$ vs. $\mbh$ relation of \cite{caixa3}.
We obtain $\mbh > 10^8$ solar masses, consistent with the reverberation measurement of $\sim 1 \times 10^9$ solar masses by \cite{fausnaugh2017}.

\begin{figure} 
	\includegraphics[width=\columnwidth]{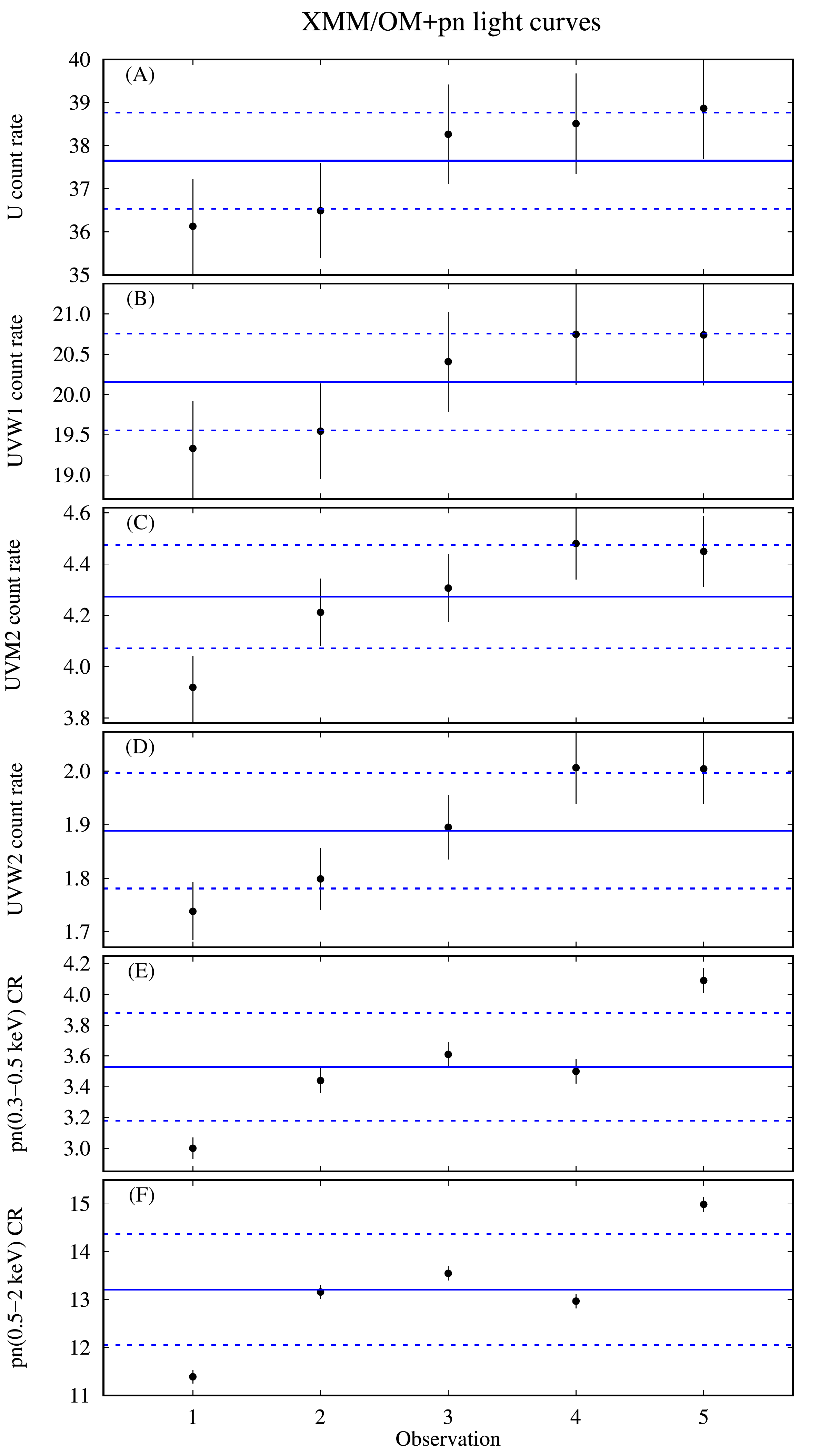}
	\caption{\label{fig:om_lc} Count rate light curves of each of the four \xmm/OM photometric filters: U (panel A), UVW1 (panel B), UVM2 (panel C), UVW2 (panel D); and count rate light curves, averaged over each observation, of \xmm/pn in the bands 0.3--0.5 keV (panel E) and 0.5--2 keV (panel F). The blue solid lines represent the mean value of the count rate over the five observations, while the blue dashed lines represent the standard deviation (i.e. the root mean square of the deviations from the mean). }
\end{figure}

\begin{figure} 
	\includegraphics[width=\columnwidth]{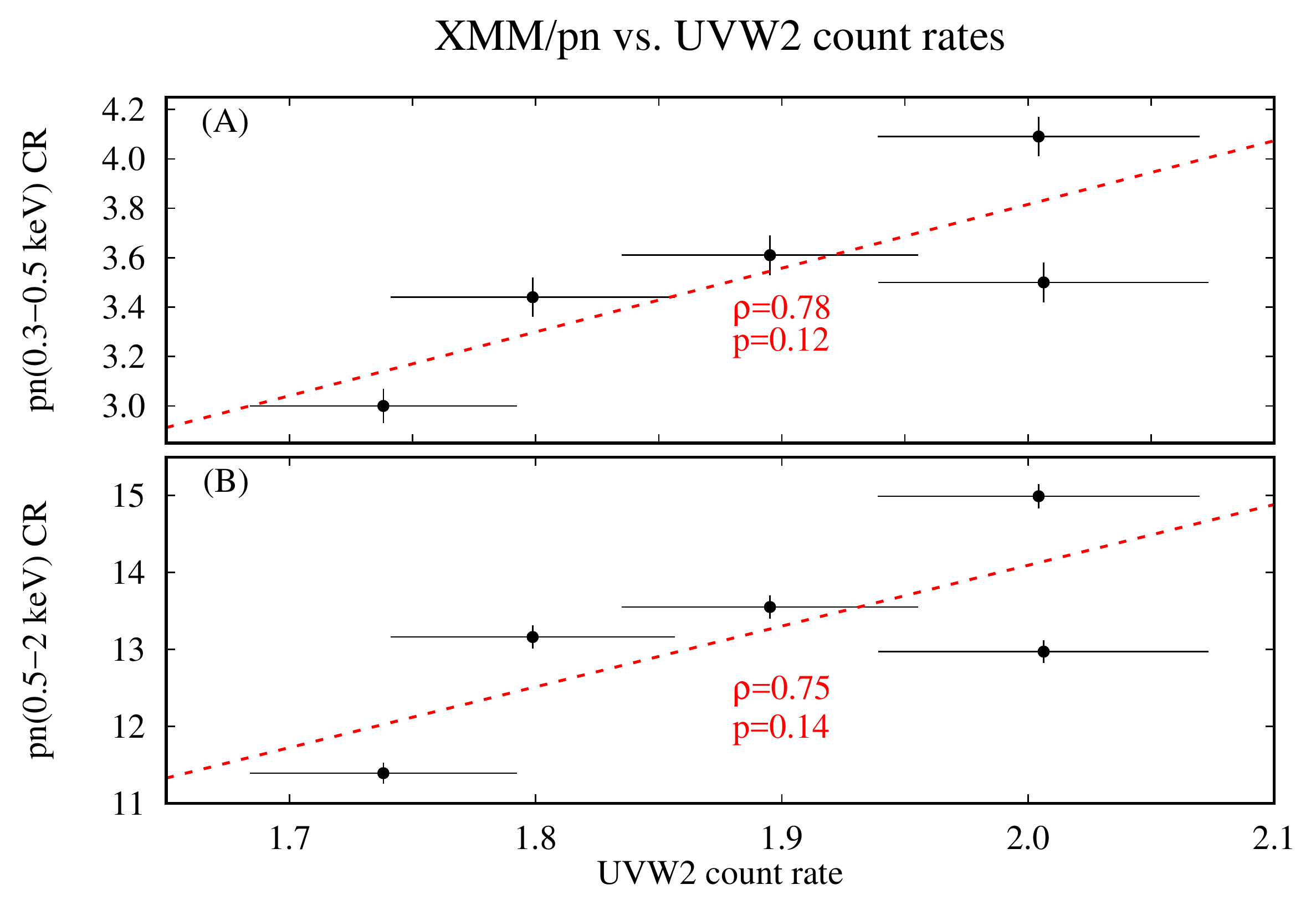}
	\caption{\label{fig:pn_vs_om} \xmm/pn count rate, averaged over each observation, in the 0.3--0.5 keV band (panel A) and in the 0.5--2 keV band (panel B) plotted against the OM/UVW2 count rate. The red dashed lines represent linear fits to the data.}
\end{figure}

In Fig. \ref{fig:om_lc} we plot the light curves of the four \xmm/OM filters, together with the \xmm/pn average count rate measured for each observation in two different bands, i.e. 0.3--0.5 keV and 0.5--2 keV. The U and UVW1 filters do not show evidence for a significant variability, while the UVM2 and UVW2 filters exhibit a variability of ${\sim} 10 \%$. In Fig. \ref{fig:pn_vs_om} we plot the \xmm/pn average count rates for each observation versus the OM/UVW2 count rate. This provides a model-independent test of a relation between the UV emission and the soft X-ray excess, which is expected to be more significant in the lower energy band. The correlation between UVW2 and 0.3--0.5 keV band has a Pearson's coefficient of 0.78, with a $p$-value of 0.12; for the 0.5--2 keV band, the Pearson's coefficient is 0.75 and the $p$-value is 0.14. Although these correlations are not highly significant, they indicate a trend of a higher X-ray flux with increasing UV flux.


\section{Spectral analysis}\label{sec:x}
We performed the spectral analysis with the \xspec\ 12.9 package \cite[][]{arnaud1996}. The RGS spectra were not binned and were analysed using the $C$-statistic \cite[][]{cstat}, to exploit the high spectral resolution of the gratings in the 0.3--2 keV band. Broad-band (UV to X-ray, 0.3--80 keV) fits were instead performed on the binned pn and \nus\ spectra plus the OM photometric data, using the $\chisq$ minimisation technique. All errors are quoted at the 90\% confidence level ($\dcash = 2.71$ or $\dchi = 2.71$) for one interesting parameter. In our fits we always included neutral absorption ({\sc phabs} model in {\sc xspec}) from Galactic hydrogen with column density $\nh = 6.98 \times 10^{20}$ \sqcm\ \cite[][]{kalberla2005}. 
We assumed the element abundances of \cite{angr} and the photoelectric absorption cross-sections of \cite{vern}.
We analysed the spectra of each observation separately. 

\begin{figure} 
	\includegraphics[width=\columnwidth,trim={0.5cm 0 2cm 1cm},clip]{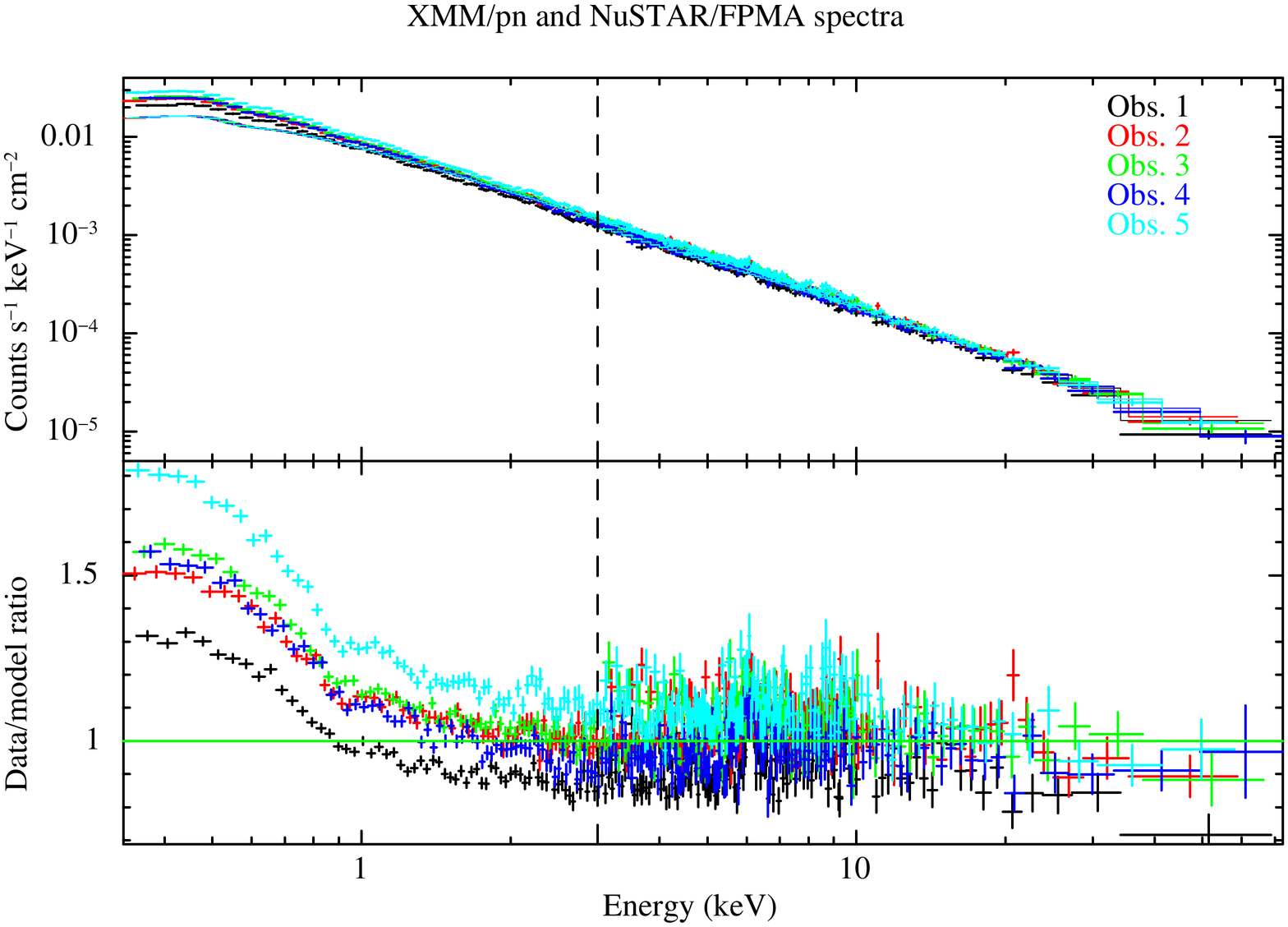}
	\caption{\label{spectra} Upper panel: the five \xmm/pn and \nus~spectra of \tc, fitted with a single power law in the 3--79 keV band. Lower panel: the ratio of the spectra to the 3--79 keV power law. Only \nus/FPMA data are shown for clarity. The data were binned for plotting purposes.}
\end{figure}

In Fig. \ref{spectra} we plot the \xmm/pn and \nus/FPMA spectra; the data were fitted in the 3--79 keV band with a simple power law with parameters tied between different detectors and observations. The extrapolation below 3 keV shows the presence of a significant soft excess. We note that \xmm/pn spectra are always flatter than \nus\ ones in the common bandpass 3--10 keV, with a difference in photon index of ${\sim}0.1$ (implying a cross-normalization of $\sim 0.7$ between pn and \nus). This discrepancy has been reported in previous \xmm/\nus\ simultaneous observations \cite[e.g.][]{cappi_5548}. In some cases, differences have been reported especially between 3 and 5 keV, where \nus\ measures a higher flux \cite[e.g.][]{fuerst2016,ponti2018}. However, in our case the spectral discrepancy is present regardless of the energy range. For this reason, in our fits we always left both the photon index and cross-normalization free to vary between pn and \nus. In the following, we report the values of the photon index and flux as measured by \nus, unless otherwise stated. The FPMA and FPMB modules are in excellent agreement with each other, with a cross-calibration factor of \ser{1.00}{0.01}.

\subsection{The RGS spectra}\label{sec:rgs}
\cite{torresi2010} found the presence of a warm absorber in \tc\ from a previous \xmm/RGS observation,
measuring a ionization parameter $\log \xi \simeq 2.7$ and a column density $\nh \simeq 3 \times 10^{22}$ \sqcm. From
the ionization parameter and the luminosity, \cite{torresi2010} estimated the location of the warm absorber to be between 10 and 60 pc. Consistently with this estimate, we did not find significant variations of the warm absorber within the observations of our campaign, even combining RGS1 and RGS2 spectra. We thus co-added the data from different epochs, separately for the two detectors RGS1 and RGS2, to obtain the best possible signal-to-noise ratio and spectral resolution. We fitted the co-added spectra in the 0.3--2 keV band. 

\begin{figure} 
	\includegraphics[width=\columnwidth]{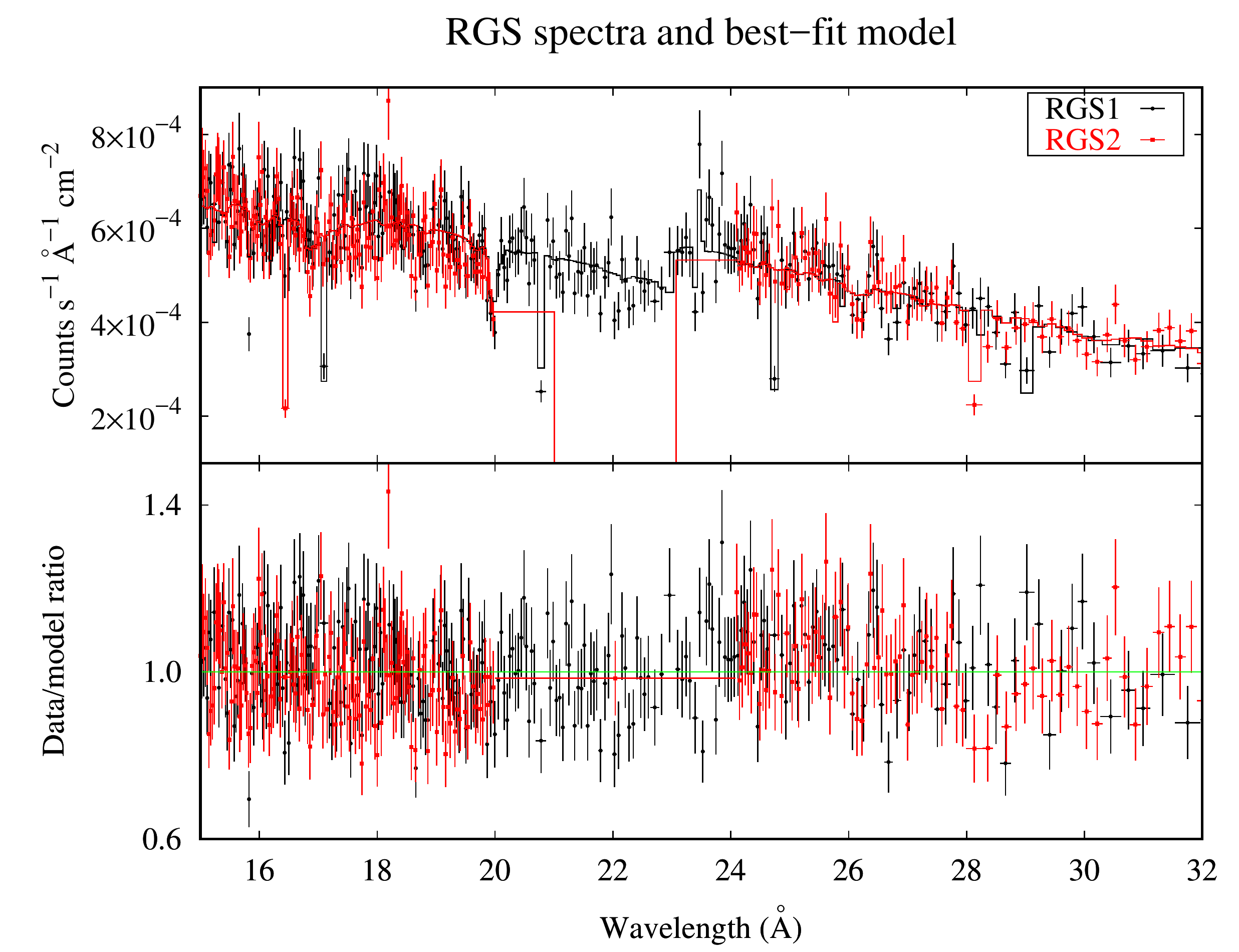}
	\caption{\label{fig:rgsfit} Upper panel: \xmm/RGS spectra (15--32 \AA) with the best-fitting model. Data are rebinned for plotting purposes. Lower panel: ratio of the spectra to the model.}
\end{figure}

First, we fitted the spectra with a simple power law, finding $\rcash=6104/5434$. Then, motivated by the results of \cite{torresi2010}, we included the ionized absorber, modelled with a table built using the spectral synthesis code \cloudy\ \cite[]{cloudy} assuming the spectral energy distribution of NGC~5548 \cite[see also][]{cappi_5548}. The allowed range for the ionization parameter $\log \xi$ is 0.1--4.9 (in units of \lumcgs\ cm), while the allowed range for the column density $\nh$ is $10^{19}{-}10^{24}$ \sqcm.
Including one absorption component (\textsc{wa1}), we found a better fit with $\rcash=5984/5430$ ($\dcash/\ddof=-120/-4$). We found a further improvement by adding a second absorption component (\textsc{wa2}), obtaining $\rcash=5935/5426$ ($\dcash/\ddof=-49/-4$) and some positive residuals around 22 \AA, that could be attributed to the K$\alpha$ emission triplet of \ovii. We then performed
a local fit at 22 \AA, on an interval 100 channels wide. In this case, given the small bandwidth, the underlying continuum is not sensitive to variations of the photon index, which was thus fixed at 2. We detected only one significant (at the 90 per cent level of confidence) emission line, which can be identified as the forbidden ($1s^2 \,^1S_0 - 1s 2s \,^3S_1$) component of the \ovii\ K$\alpha$ triplet. Indeed, the energy of this line was found to be \ser{559}{3} eV (or \ser{22.18}{0.12} \AA), while the theoretical energy of the forbidden \ovii\ line is 561 eV \cite[or 22.10 \AA;][]{atomdb}. The line flux was found to be \aerexp{3}{2}{1}{-5} photons \sqcm\ \pers. Including this line in the fit over the 0.3--2 keV band, we found $\rcash=5921/5424$ ($\dcash/\ddof=-14/-2$) and no further residuals that can be attributed to strong atomic transitions. Finally, allowing the covering fraction of the warm absorber components to vary, we found it to be consistent with unity. The RGS spectra with best-fitting model are plotted in Fig. \ref{fig:rgsfit}, while the best-fitting parameters of the warm absorber are reported in Table \ref{tab:rgs}. The two components have different outflow velocities $v$, up to $-2400$ \kms\ for the higher ionization phase. The results are in rough agreement with those found by \cite{torresi2010}
%, namely $\log \xi \simeq 2.7$ and $\nh \simeq 3 \times 10^{22}$ \sqcm\ 
for a single-zone absorber (see also Sect. \ref{sec:broad}).

\begin{table}
	\begin{center}
		\caption{Best-fitting parameters of the warm absorbers (\textsc{wa1, wa2}) for the RGS spectra. 
			%$\sigma_v$ is the turbulent velocity, 
			$v$ is the velocity shift with respect to the systemic velocity of \tc. \label{tab:rgs}}
		\begin{tabular}{l c}
			\hline 
			$\log \xione $ (\lumcgs\ cm) & \ser{ 3.07 }{ 0.09 } 
			%\\  
			%$\log \vturbone$ (\kms) & Unconstr.
			\\  $\log \nhone$ (\sqcm) & \aer{ 21.72 }{ 0.04}{ 0.02 } 
			\\ $\vone$ (\kms) & \ser{ -2400 }{ 300 } 
			\\  
			%\hline 
			\noalign{\medskip} 
			$\log \xitwo$ (\lumcgs\ cm) & \ser{ 2.1 }{ 0.1 }
			%\\ $\log \vturbtwo$ (\kms) & Unconstr.
			\\  $\log \nhtwo$ (\sqcm)& \ser{ 21.5 }{ 0.1 } 
			\\ $\vtwo$ (\kms)& \ser{-300}{300}
			\\ 
			\hline 
		\end{tabular}
	\end{center}
\end{table}

\subsection{The Fe K\boldmath{$\alpha$} line}\label{sec:line}
\begin{figure} 
	\includegraphics[width=\columnwidth,trim={0.5cm 0 2cm 0},clip]{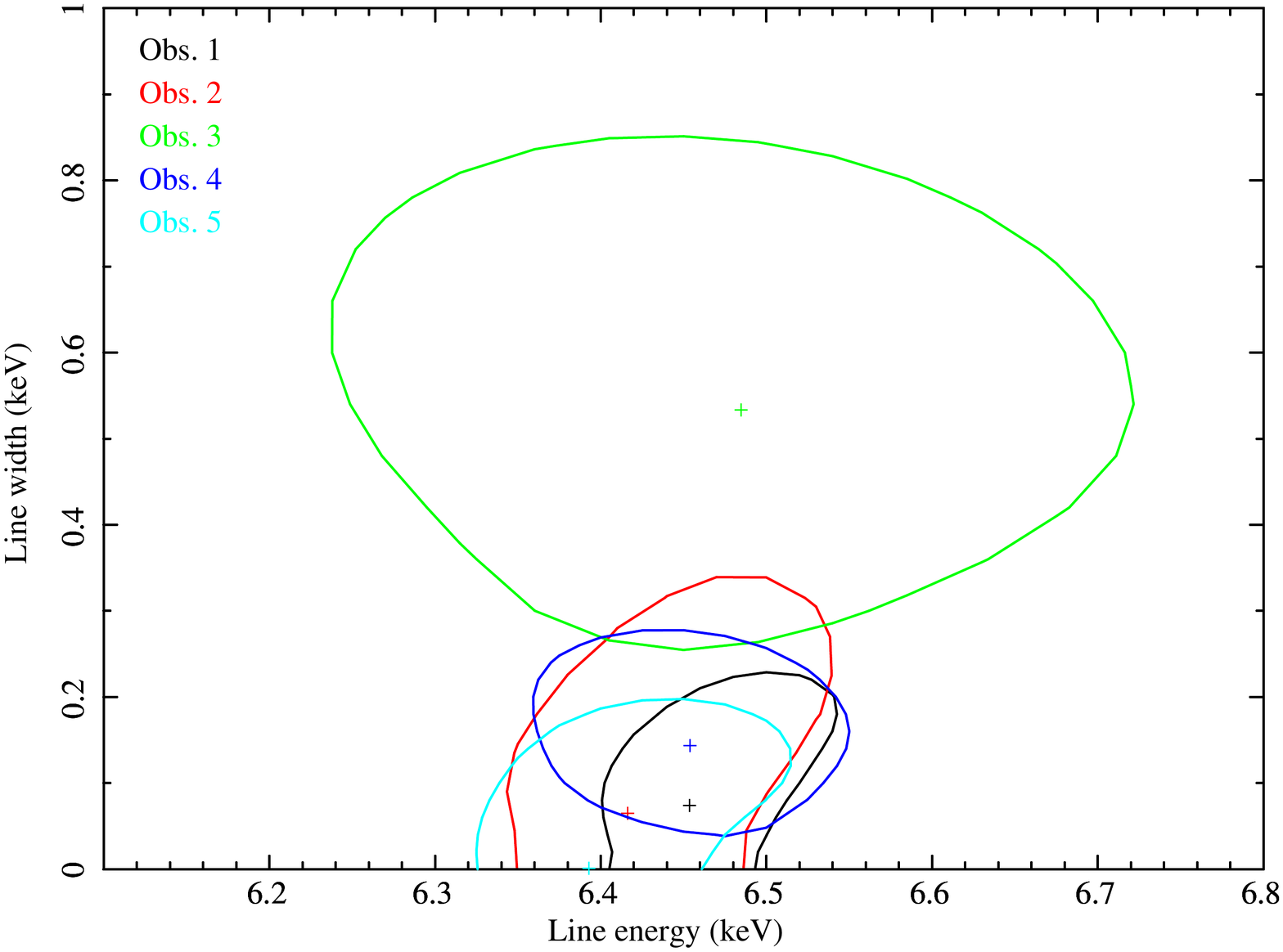}
	\caption{\label{fig:cont_en_sigma} Contour plots of the \fek\ line intrinsic width versus rest-frame energy at the 90\% confidence level.}
\end{figure}

\begin{figure} 
	\includegraphics[width=\columnwidth,trim={0.5cm 0 2cm 0},clip]{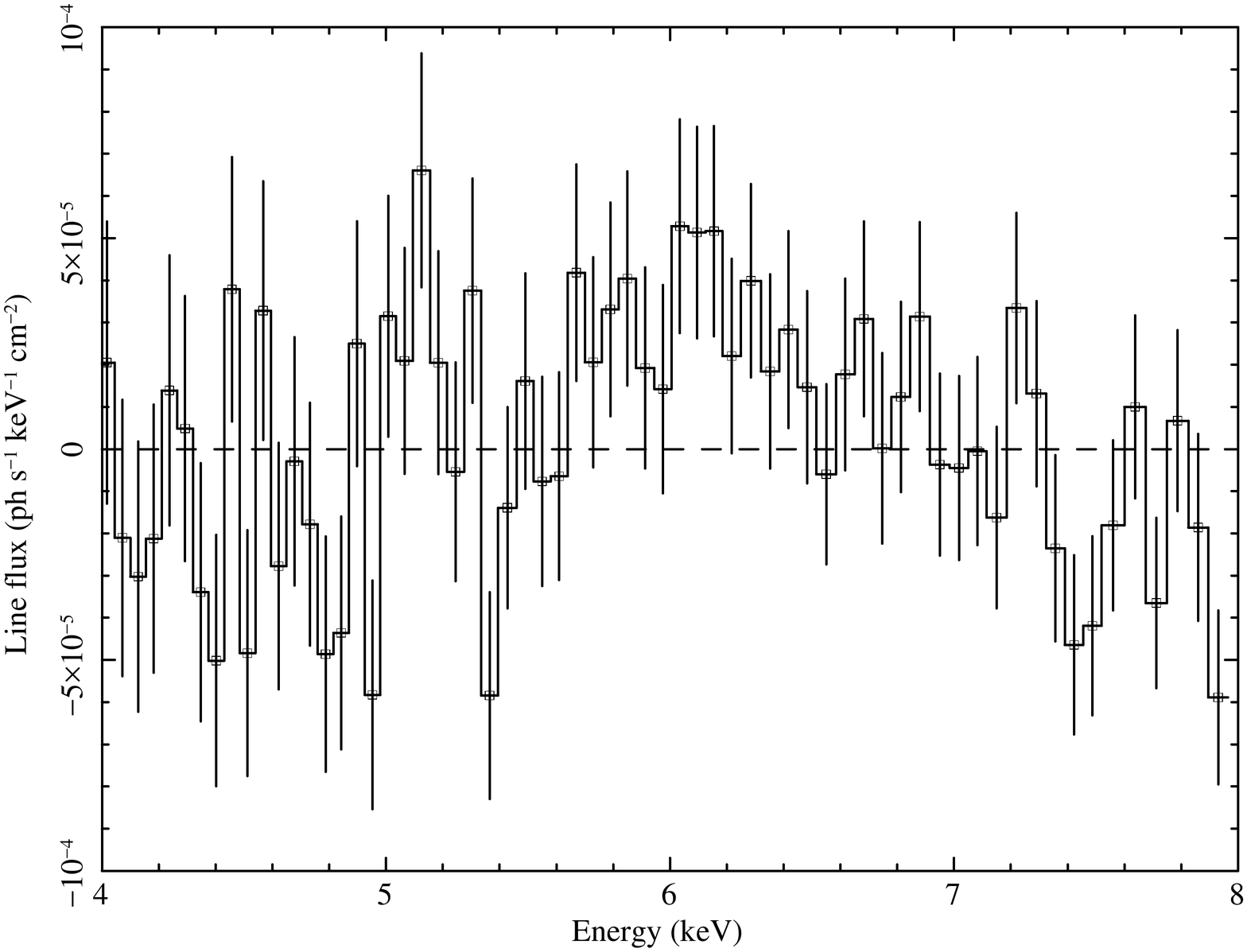}
	\caption{\label{fig:line_res} \fek\ line profile in observation 3, from \xmm/pn data (observer's frame). The plot shows residuals of a fit in the 3--10 keV band including a simple power law.}
\end{figure}

\begin{figure}
	\includegraphics[width=\columnwidth]{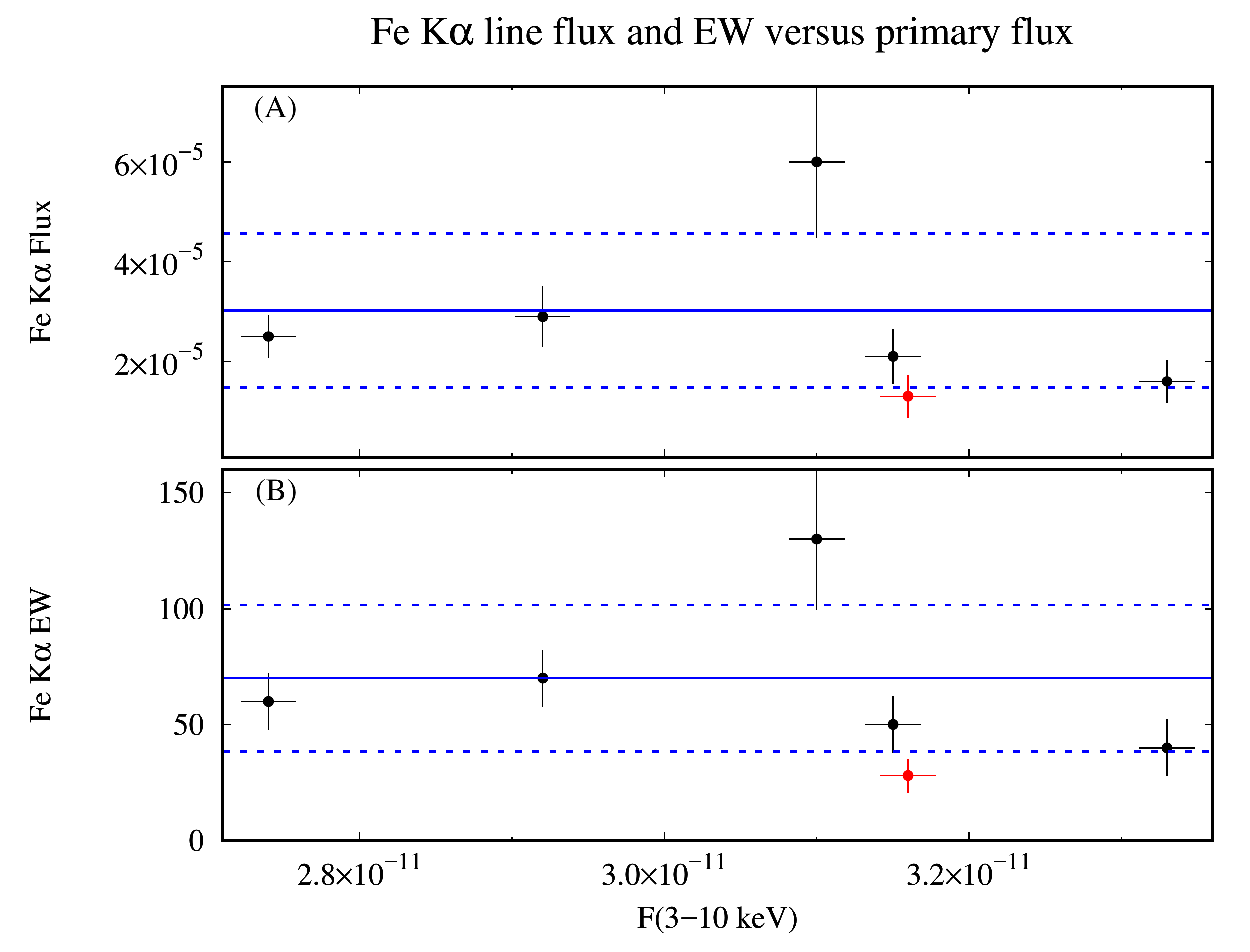}
	\caption{Parameters of the \fek\ line plotted against the primary flux in the 3--10 keV band. The red point corresponds to the 2008 \xmm/pn spectrum.
		Panel (A): the line flux in units of %$10^{-5}$
		photons \sqcm~\pers. 
		Panel (B): the line equivalent width (EW) in units of eV. 
		Error bars denote the $1\sigma$ uncertainty. The blue solid lines represent the mean value for each parameter during the campaign, while the blue dashed lines represent the standard deviation. }
	\label{fig:line_flux}
\end{figure}

To investigate the shape and variability of the \fek\ line at 6.4 keV, we used \xmm/pn data between 3 and 10 keV, given the superior energy resolution and throughput compared with \nus\ in that energy band. We simultaneously fitted the five pn spectra with a model including a variable power law plus a Gaussian line with a variable flux.
We first assumed an intrinsically narrow line, i.e. the intrinsic width $\sigma$ was fixed at zero,  finding $\rchisq= 536/515$.
Next, we left $\sigma$ free but tied between the different observations, finding $\rchisq=528/514$ (i.e. $\dchi/\ddof = -8/{-}1$) and $\sigma=\aerm{0.11}{0.6}{0.5}$ keV.
Then we left $\sigma$ free to vary between the different observations, obtaining a better fit ($\rchisq=513/510$, i.e. $\dchi/\ddof = -15/{-}4$). The improvement is mostly due to the broadening of the line in observation 3, where $\sigma=\aerm{0.5}{0.3}{0.2}$ keV; the contours of the line intrinsic width versus rest-frame energy are plotted in Fig. \ref{fig:cont_en_sigma}. Indeed, considering observation 3 alone, the zero-width Gaussian line fit gives $\rchisq= 125/104$ while the free-width Gaussian line fit gives $\rchisq= 109/103$ ($\dchi/\ddof = -16/{-}1$ and probability of chance improvement less than \expo{2}{-4} from an F-test). We found no improvement by adding a second, narrow line component in observation 3, neither at 6.4 keV nor at higher energies (such as the K$\alpha$ lines of Fe \textsc{xxv} at 6.7 keV or Fe \textsc{xxvi} at 6.966 keV). 
The properties of the \fek\ line are summarized in Table \ref{tab:line}.
We plot in Fig. \ref{fig:line_res} the profile of the \fek\ line in observation 3. Fig. \ref{fig:line_res} also shows a hint of an absorption feature at around 7.4 keV (i.e. $\sim 7.8$ keV rest-frame). However, adding a narrow Gaussian absorption component only marginally improves the fit ($\rchisq= 104/101$, i.e. $\dchi/\ddof = -5/{-}2$ and probability of chance improvement of 0.09).
Also in observation 4 we found a marginal evidence for a line broadening ($\sigma=\aerm{0.13}{0.09}{0.08}$ keV), however the improvement is less significant in this case (for the single spectrum, $\dchi/\ddof = -5/{-}1$ and probability of chance improvement of 0.03).
We also note that, at least in observation 1, the \fek\ line is blueshifted by $\sim 50$ eV with respect to the theoretical value of 6.4 keV. This could be due to a known calibration problem in pn data, an effect of the long-term degradation of the EPIC/pn charge transfer inefficiency (CTI). The shift is present using either single plus double or single-only events, despite the use of the latest correction files for CTI and procedure as described in Smith et al. (2014, \textsc{xmm-ccf-rel-323}\footnote{\url{http://xmm2.esac.esa.int/docs/documents/CAL-SRN-0323-1-1.ps.gz}}). Moreover, the energy and width of the line are poorly constrained with MOS or \nus\ data. To correct for this uncertainty, we fixed the Gaussian line energy at 6.4 keV, i.e. assuming production by ``cold'' iron (less ionized than Fe \textsc{xii}), leaving the redshift free to vary in pn data.
We also re-analysed the archival 2008 \xmm/pn spectrum, to gain some further insight into the properties and the temporal evolution of the \fek\ line. Fitting the spectrum as above, we found the presence of a line with an energy of \aer{6.36}{0.08}{0.06} keV, an intrinsic width $<0.2$ keV and an equivalent width of \ser{28}{12} eV.
We plot in Fig. \ref{fig:line_flux} the flux and equivalent width of the \fek\ line against the primary flux in the 3--10 keV band for our observations and that of 2008 (red point). No prominent variability is observed, although we have a hint of a higher line flux in observation 3.
\begin{table}
	\begin{center}
		\caption{
			The properties of the \fek\ emission line. $E$ is the energy of the line (rest-frame) in keV, $\sigma$ is the intrinsic line width in keV, the flux is in units of $10^{-5}$ photons \sqcm\ \pers and EW is the equivalent width in eV.
			\label{tab:line}}
		\begin{tabular}{l c c c c c}
			\hline 
			&Obs. 1&Obs. 2&Obs. 3&Obs. 4&Obs. 5\\
			\hline
			$E$&\aer{6.45}{0.09}{0.04}&\aer{6.42}{0.12}{0.07}&\ser{6.5}{0.2}&\ser{6.45}{0.09}&\aer{6.40}{0.10}{0.08}\\
			$\sigma$&$<0.2$&$<0.35$&\aer{0.5}{0.3}{0.2}&\aer{0.13}{0.09}{0.08}&$<0.2$\\
			flux &\ser{2.6}{0.7}&\ser{2.2}{0.8}&\ser{4.6}{1.6}&\ser{2.7}{0.8}&\ser{1.8}{0.8}\\
			EW &\ser{60}{20}&\ser{50}{20}&\ser{120}{40}&\ser{60}{20}&\ser{40}{20}\\  
			\hline
		\end{tabular}
	\end{center}
\end{table}

If the \fek\ line is produced in the accretion disc, the broad profile could be due to relativistic blurring. Therefore, 
we tested a model including a narrow ($\sigma=0$) Gaussian line blurred by relativistic effects from the accretion disc (\kdblur\ convolution model in \xspec).  We left the inner disc radius $\rin$ free to vary between the different observations, fixing the outer disc radius to 400 \rg\ (as the fit is insensitive to this parameter). The disc inclination was left free, but tied between the different observations. We obtained a fit with $\rchisq=519/509$, i.e. slightly worse than the one including the simple Gaussian line ($\dchi/\ddof = +6/{-}1$). We found an upper limit of 23 deg to the inclination, while the inner disc radius was found to be \aer{8}{20}{4} \rg\ in observation 3, and unconstrained in the other observations. These results might suggest a variation of the inner radius of the disc, which could be truncated in the inner region except during observation 3. On the other hand, the lack of strong variability of the line across the campaign might rather suggest an origin from material lying at least 1-2 light months away from the nucleus. We further discuss this point in the next sections.
% chisq/dof = 112/106 vs 95/103

\subsection{A reflection component?}\label{sec:reflection}
We investigated the presence of a reflection component associated with the \fek\ line fitting the pn and \nus\ data in the 3--79 keV energy band. Fitting the spectra above 3 keV allows us to focus on the putative Compton hump avoiding the complexities in the soft band (soft excess and warm absorber).

First, we fitted the five pn+\nus\ data sets with a simple model including a power law with an exponential cut-off plus a Gaussian line (see Sect. \ref{sec:line}). We left the photon index, cut-off energy and normalization of the power law free to vary between the different observations. Concerning the Gaussian line, we left the width and flux free to vary between the different observations
. We found a good fit ($\rchisq=1925/1921$), and only lower limits to the high-energy cut-off of 130--150 keV, consistent with the measurements reported by \cite{grandi2001} and \cite{ballantyne2014}. 
Then, to test for the presence of a reflection continuum, we replaced the cut-off power law with the \pexrav\ model in \xspec, which includes Compton reflection from a neutral medium of infinite column density in a slab geometry \cite[]{pexrav}, such as a standard neutral accretion disc. We fixed the inclination angle at 30 deg, since the fit was not sensitive to this parameter. We found no improvement, and only an upper limit to the reflection fraction $\mathcal{R} = \Omega / 2 \pi$, where $\Omega$ is the solid angle subtended by the reflector, of ${\sim} 0.1$. To further test the robustness of this result, we performed the same fit to \nus\ data alone, thus avoiding cross-calibration issues. We found the same upper limit to the reflection fraction.

Next, we tested a model including \relxill, version 1.0\footnote{\url{http://www.sternwarte.uni-erlangen.de/~dauser/research/relxill/}}, which describes relativistically blurred reflection off an ionized accretion disc \cite[][]{relxill,dauser2016}. We thus replaced the cut-off power law and Gaussian line with \relxill, which self-consistently incorporates the Compton hump and the fluorescence lines. We left free to vary between the different observations the inner disc radius, which determines the broadening of the \fek\ line. The reflection fraction $\mathcal{R}$, inclination $i$, ionization parameter $\xi$ and iron abundance $\afe$ were also free parameters, but tied among the different observations since they were found to be consistent with being constant. We found a good fit ($\rchisq=1940/1923$) with $\mathcal{R}=\serm{0.08}{0.02}$, $i<15$ deg, $\log \xi <0.7$, $\afe>7$, and only lower limits to the inner disc radius of 20--40 \rg. Similar constraints for \relxill\ were found from the analysis of previous \nus\ data by \cite{ballantyne2014}. According to these results, relating the \fek\ line to a reflection component from the disc would require a large iron overabundance and the disc to subtend a small angle $\Omega= 2 \pi \mathcal{R} \simeq 0.16 \pi$. 

The lack of a significant reflection continuum is consistent with the results of \cite{ballantyne2014}, and might indicate that the \fek\ line originates from
Compton-thin material ($\nh = 10^{22} {-} 10^{23}$ \sqcm), which does not produce a prominent Compton hump. To test this hypothesis, we replaced the \pexrav\ component and the Gaussian line with the \mytorus\ model, which includes Compton reflection and iron fluorescent lines from a gas torus with an opening angle of 60 deg \cite[]{mytorus,mytorus2}. We fixed the inclination angle of the torus at 30 deg, with no improvement by leaving it free to vary. The column densities of the scattered and line components were linked and free to vary, but tied between the different observations as they were consistent with each other. The photon indices and normalizations of the scattered and line components were tied to those of the primary power law, allowing for a relative normalization by means of a multiplicative constant. 
We assumed the standard \mytorus\ configuration $A_S=A_L$, where $A_S$ is the scaling factor for the scattered component and $A_L$ that of the line component \cite[][]{mytorus2}. 
As a first step, we assumed $A_S=1$, which corresponds to a covering fraction of 0.5. 
Fitting the data with this model, we found a worse fit than the one including a simple cut-off power law plus Gaussian line ($\rchisq=1957/1930$, i.e. $\dchi/\ddof={+}32/{+}9$). Next, we left $A_S$ free, finding a statistical improvement ($\rchisq=1949/1929$, i.e. $\dchi/\ddof={-}8/{-}1$).
We found a further improvement ($\rchisq=1936/1924$, i.e. $\dchi/\ddof={-}13/{-}5$ and $\dchi/\ddof={+}11/{+}3$ with respect to the power law plus Gaussian) by smoothing the line component of \mytorus\ with a Gaussian profile (\textsc{gsmooth} model in \xspec). The energy width $\sigma$ is consistent with zero except in observation 3, where $\sigma=\aerm{0.22}{0.15}{0.10}$ keV.
We obtained $A_S=\aerm{0.45}{0.05}{0.08}$ and a column density of \serexp{6}{2}{23} \sqcm. 

The results obtained with \pexrav, \relxill\ and \mytorus\ are summarized in Table \ref{tab:refl}.
\begin{table}
	\begin{center}
		\caption{
			Best-fitting parameters, common to all the observations, of the three reflection models tested using the data above 3 keV. The inclination $i$ is in deg, $\xi$ is in units of \lumcgs\ cm, $\nh$ is in \sqcm. Parameters in italics were frozen because of poor constraints.
			\label{tab:refl}}
		\begin{tabular}{l c c c }
			\hline 
			&\pexrav&\relxill&\mytorus \\
			\hline
			$\mathcal{R}$&\lsup{0.1}&\ser{0.08}{0.02}&-\\
			$i$&\textit{30}&\lsup{15}&\textit{30}\\
			$\log \xi$&-&\lsup{0.7}&-\\
			$\afe$&\textit{1}&\linf{7}&-\\
			$A_S$&-&-&\aer{0.45}{0.05}{0.08}\\
			$\nh$&-&-&\serexp{6}{2}{23}\\
			$\rchisq$&1925/1920&1940/1923&1936/1924\\  
			\hline
		\end{tabular}
	\end{center}
\end{table}

\subsection{The broad-band fits}\label{sec:broad}
\begin{figure} 
	\includegraphics[width=\columnwidth,trim={0.5cm 0 2cm 1cm},clip]{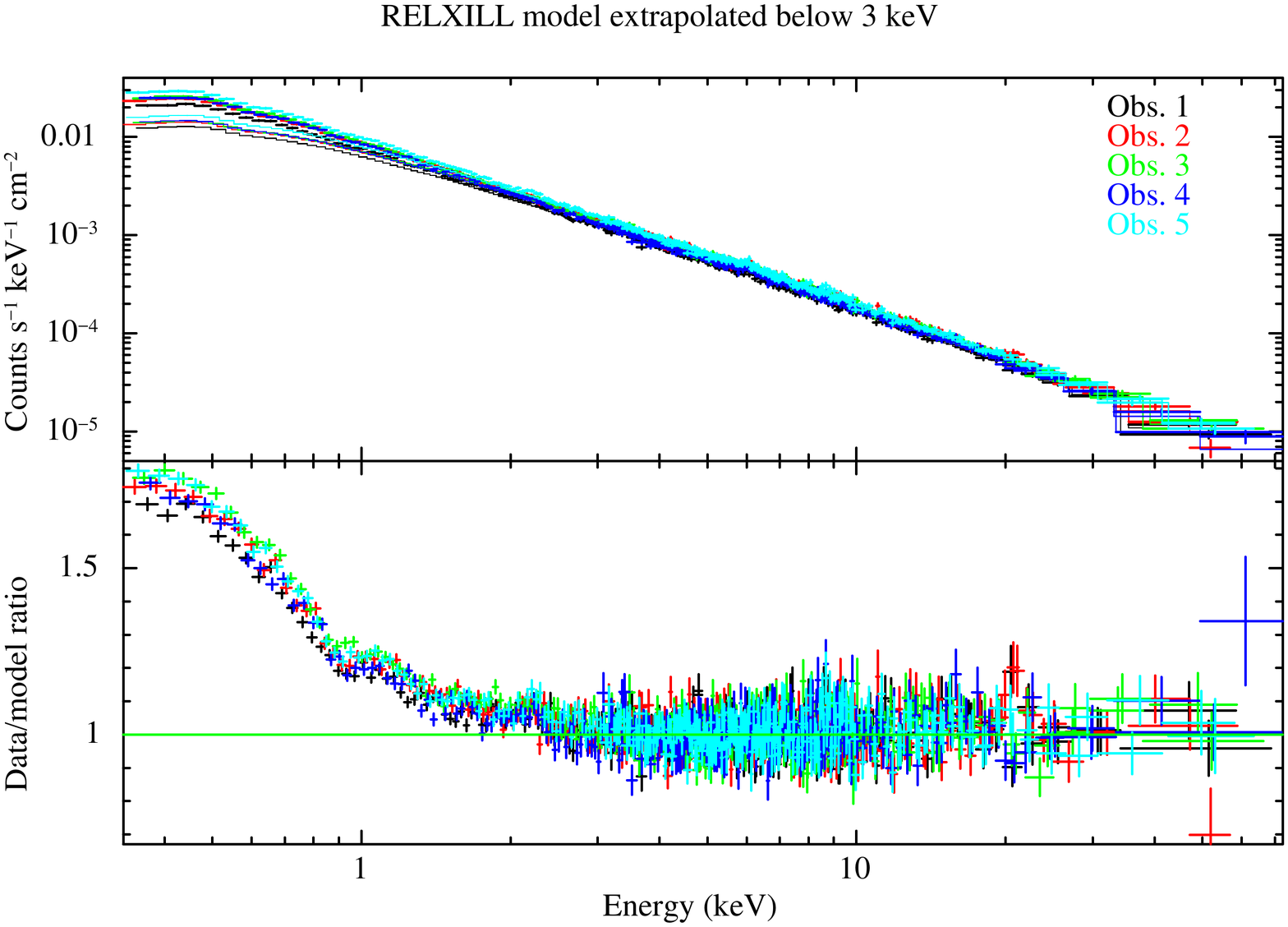}
	\caption{\label{fig:relxill} Upper panel: pn and \nus~spectra fitted with \relxill\ in the 3--79 keV band (see Sect. \ref{sec:reflection}). Lower panel: the ratio of the broad-band (down to 0.3 keV) spectra to the model. The data were binned for plotting purposes.}
\end{figure}

As the final step of the high-energy spectral analysis, we fitted the pn and \nus\ data in the whole X-ray energy band (0.3--79 keV) including the optical/UV data from the OM.
First, we note that the extrapolation of the best-fitting model above 3 keV to lower energies clearly reveals a soft excess, even in the case of the ionized reflection model with \relxill\ (see Fig. \ref{fig:relxill}). Even refitting the data in the 0.3--79 keV with \relxill, and including the warm absorber (Sect. \ref{sec:rgs}), we obtained a poor fit ($\rchisq=2712/2249$) with significant residuals below 2 keV. 

The broad-band (UV to hard X-rays) data allow us to test physical Comptonization models. The different components are described below.
\subsubsection{The primary continuum and soft excess}
The hard X-ray spectrum is modelled with \nthcomp\ \cite[]{nthcomp1,nthcomp2}, which describes thermal Comptonization. The main parameters of \nthcomp\ are the electron temperature $\kte$, the seed photon temperature $\ktbb$ \cite[we assumed a multicolour disc blackbody distribution:][]{mitsuda1984,makishima1986}, and the photon index $\Gamma$ of the asymptotic power law.
The soft excess in this source is not correctly reproduced by ionized reflection, also given the lack of a significant Compton hump. We thus tested a warm Comptonization scenario for the soft excess, which was also modelled with \nthcomp. Therefore, our model included a hot \nthcomp\ component characterized by an electron temperature $\kteh$ and a photon index $\gammah$, and a warm \nthcomp\ component characterized by $\ktew$ and $\gammaw$ (see Table \ref{params}). We assumed the same seed temperature $\ktbb$ for both (warm/hot) components. 
\subsubsection{Absorption}
We included a warm absorber, modelled with \cloudy, following Sect. \ref{sec:rgs}. In the case of pn, a single warm absorber component was found to adequately fit the data, possibly because of the lower spectral resolution compared with RGS. Adding a second absorber did not improve the fit. To account for cross-calibration uncertainties between pn and RGS, especially below 0.5 keV, the parameters of the absorber were free, but tied among different observations (with no significant improvement by leaving them free to change). 
Since we used optical/UV data, we also included interstellar extinction (\redden\ model in \xspec) with $E(B-V)=0.0598$ \cite[]{schlegel1998}. 
\subsubsection{Emission lines}
We included a Gaussian line to account for the \fek\ line at 6.4 keV (Sect. \ref{sec:line}). Since the line is not accompanied by a significant Compton hump (Sect. \ref{sec:reflection}), we assumed no reflection component. 
%\cite[see also][]{ballantyne2014}. 
We also included another Gaussian line to account for residuals around 0.5 keV, most likely corresponding to the \ovii\ line at 0.56 keV found from the analysis of RGS data (Sect. \ref{sec:rgs}).
\subsubsection{Small blue bump}
The small blue bump is a broad feature generally seen in the optical/UV spectrum of AGNs between 2000--4000 \AA, due to a blend of strong \feii\ lines and the Balmer continuum emission \cite[e.g.][]{grandi1982,wills1985}. To take into account this component, we produced a table model for \xspec\ (\smallbb) using the calculations of \cite{wills1985} and \cite{grandi1982} for the \feii\ lines and for the Balmer continuum, respectively \cite[for a detailed discussion, see also][]{mehdipour}. 
%The only free parameter of this component was its normalization. 
From the fit, the flux contribution of this component was found to be \serexp{6.2}{0.5}{-12} \fluxcgs, and consistent with being constant among the different observations.
\begin{table*}
	\begin{center}
		\caption{Best-fitting parameters of the broad-band model described in Sect. \ref{sec:broad}: {\sc WA*(smallBB+nthcomp,w+nthcomp,h+zgauss1+zgauss2)} in \xspec\ notation. In the second column we report the fit parameters that were linked among all observations. In the subsequent columns, we report the fit parameters that were free to vary for each observation. The parameters of the Gaussian line at 6.4 keV (\textsc{zgauss1}) are reported in Table \ref{tab:line}.  \label{params}
		}
		\begin{tabular}{l c c c c c c}
			\hline & all obs. & obs. 1 & obs. 2 & obs. 3 & obs. 4 & obs. 5 \\
			\hline
			$F_{\textsc{smallBB}}$ ($10^{-12}$ \fluxcgs)& \ser{6.2}{0.5} &&&&& \\ 
			\noalign{\medskip} 
			$\log \xi$ (\lumcgs\ cm) & \ser{2.87}{0.04} &&&&& \\
			$\log \vturb $ (km \pers) & \aer{1.82}{0.09}{0.05} &&&&&\\
			$\log \nh$ (\sqcm) & \ser{22.15}{0.07} &&&&& \\
			%\hline 
			\noalign{\medskip} 
			$\gammaw$ && \ser{ 2.54 }{ 0.05 }&\ser{ 2.50 }{ 0.05 }& \ser{ 2.44 }{ 0.05} & \ser{ 2.49}{ 0.04 }& \ser{ 2.39}{ 0.04}\\
			$\ktew$ (keV) && \aer{ 0.5 }{ 0.2 }{ 0.1 } & \ser{ 0.6 }{ 0.1 } & \aer{ 0.7}{ 0.2 }{ 0.1 }  & \ser{ 0.5 }{ 0.1 }  & \aer{ 0.7 }{ 0.2 }{ 0.1 }\\
			$N_{\textsc{nthcomp,w}}$ ($10^{-3}$)&& \aer{ 1.9 }{ 0.6 }{ 0.4 } & \aer{ 2.4 }{ 0.7 }{ 0.5 }  & \aer{ 3.7 }{ 0.8 }{ 0.7 }  & \aer{ 2.8 }{ 0.6 }{ 0.5 }  & \ser{ 5 }{1 }\\
			$\ktbb$ (eV) & \aer{3.4}{0.4}{0.2} &&&&& \\
			\noalign{\medskip} 
			$\gammah$ && \ser{ 1.78 }{ 0.01 }&\ser{ 1.79 }{ 0.01 }& \ser{ 1.78 }{ 0.02} & \ser{ 1.78}{ 0.01 }& \ser{ 1.79}{ 0.02}\\
			$\kteh$ (keV) && \linf{20} &\linf{20}&\linf{40}&\linf{30}&\linf{40} \\
			$N_{\textsc{nthcomp,h}}$ ($10^{-2}$)&& \ser{ 1.04 }{ 0.03 } & \ser{ 1.21 }{ 0.04 } & \aer{ 1.18 }{ 0.04 }{ 0.06 }  & \aer{ 1.09 }{ 0.04 }{ 0.02 }  & \aer{ 1.23 }{ 0.03 }{0.08}\\
			\noalign{\medskip} 
			$E_{\textsc{zgauss2}}$ (keV) &\ser{0.59}{0.02}&&&&& \\ 
			$N_{\textsc{zgauss2}}$ ($10^{-4}$) &\ser{1.3}{0.4}&&&&& \\ 
			\noalign{\medskip} 
			$\ftwoten$ ($10^{-11}$ \fluxcgs) &&3.4&3.9& 3.9&3.7&4.2 \\
			$\lbol$ ($10^{45}$ \lumcgs) &&2.6&2.8& 3.4&2.9&3.5 \\			
			\noalign{\medskip} 
			$\rchisq$ &$2430/2254$&&&&&\\
			\hline			
		\end{tabular}
	\end{center}
\end{table*}
%\\[\baselineskip]
\subsubsection{Results}
\begin{figure*} 
	\includegraphics[width=\textwidth]{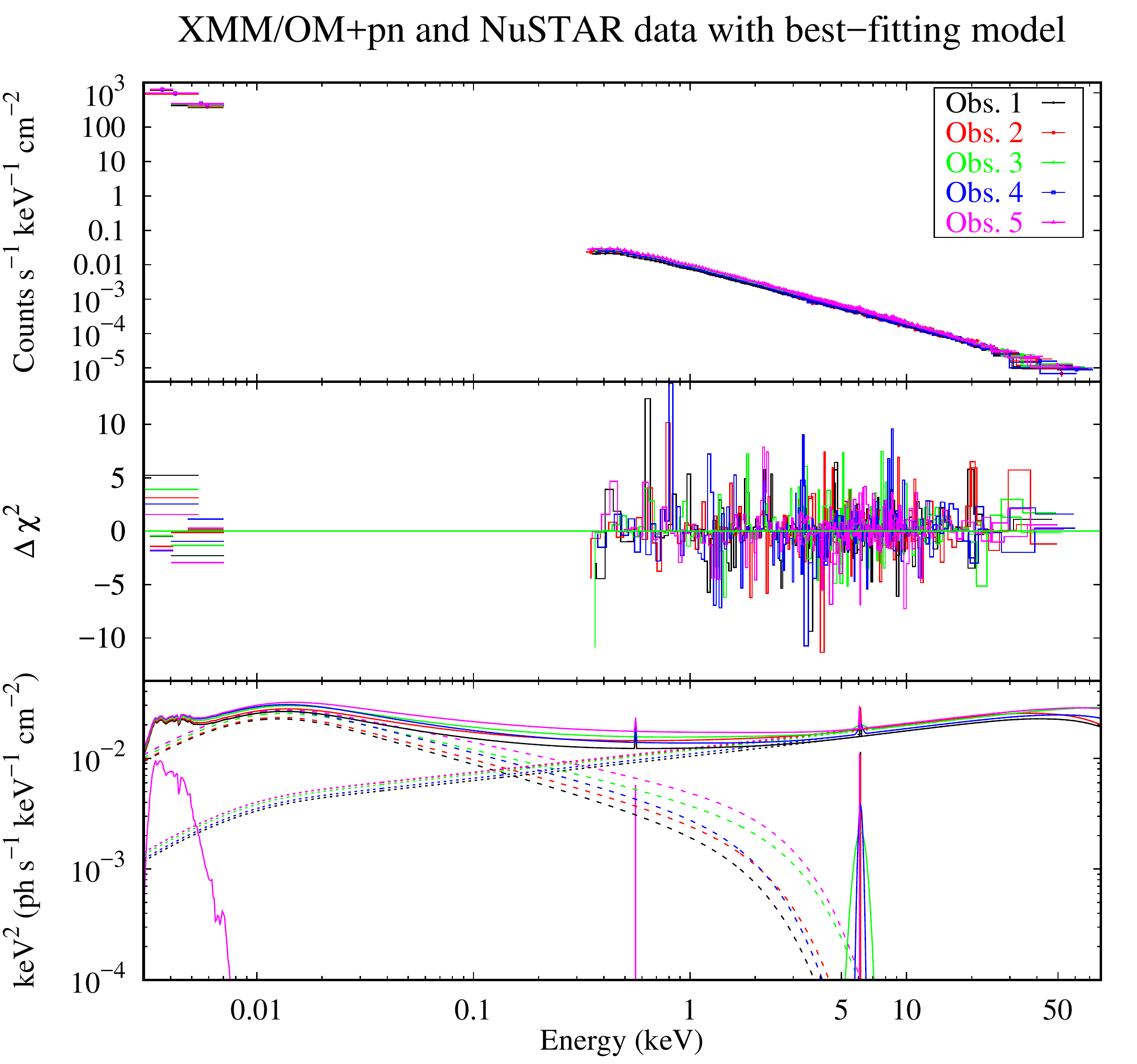}
	\caption{\label{fig:fit} Broad-band UV/X-ray data and best-fitting model (see Table \ref{params}). Upper panel: \xmm/OM, pn and \nus\ data (rebinned for plotting purposes) with folded model. Middle panel: contribution to $\chi^2$. Lower panel: best-fitting model $E^2 f(E)$, without any absorption, with the plot of the warm and hot \nthcomp\ components (dashed and dotted lines, respectively), the small blue bump and Gaussian lines at 0.59 keV and 6.4 keV (solid lines).}
\end{figure*}

\begin{figure} 
	\includegraphics[width=\columnwidth]{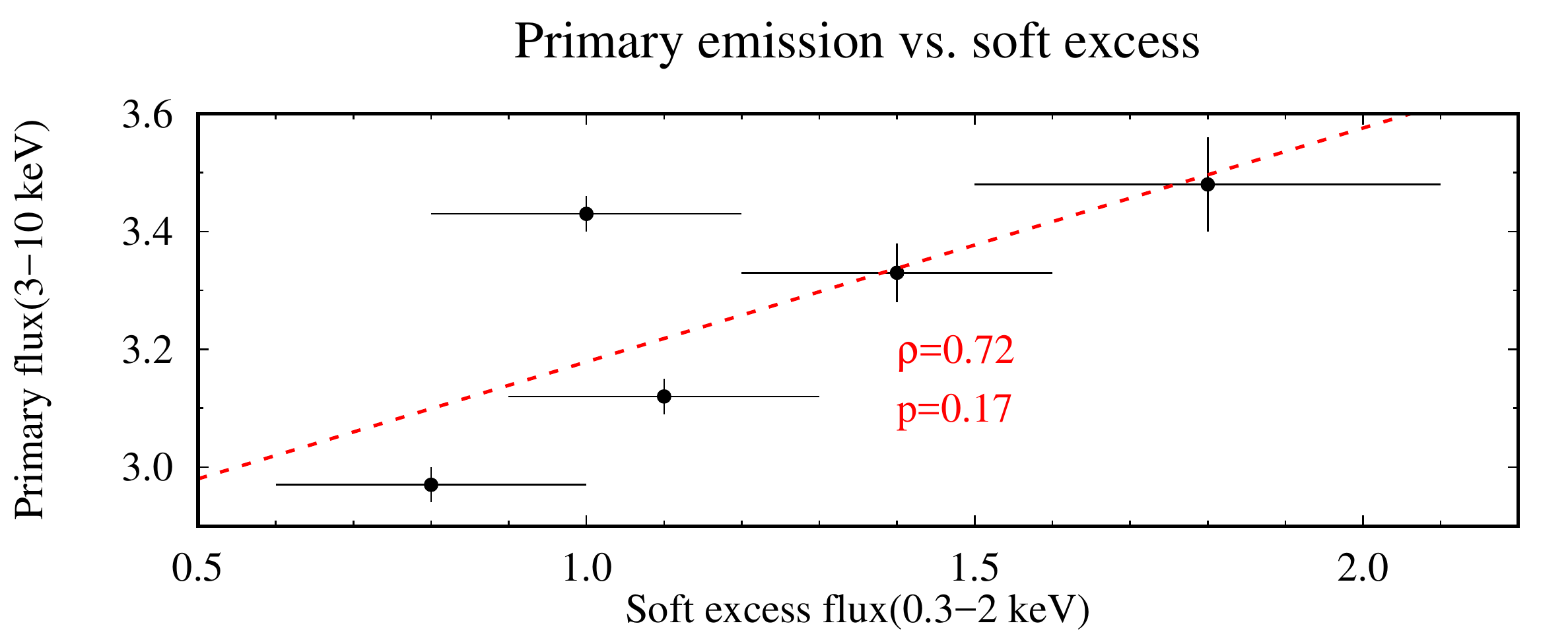}
	\caption{\label{fig:fluxes_wh} The primary X-ray flux (3--10 keV) of the hot \nthcomp\ component versus the flux (0.3--2 keV) of the warm \nthcomp\ component, as measured in the different observations. The fluxes are in units of $10^{-11}$ \fluxcgs. The red dashed line represents a linear fit to the data.}
\end{figure}

In Fig. \ref{fig:fit} we show the data, residuals and best-fitting model, while all the best-fitting parameters are reported in Table \ref{params}.
The parameters of the single-zone warm absorber are consistent with those found by \cite{torresi2010}.
For the warm \nthcomp\ component, we find a photon index in the range 2.4--2.5 and a temperature always consistent with 0.6 keV. The corresponding optical depth, as derived from the \nthcomp\ model, is around 20. 
For the hot \nthcomp\ component, we find a photon index of around 1.8 and only a lower limit to the temperature of 20--40 keV. This corresponds to an upper limit to the optical depth of $\sim 4$.
The absorption-corrected model luminosities are found to be $2.6{-}3.5 \times 10^{45}$ \lumcgs\ for the total spectra (0.001--100 keV). Since the Eddington luminosity for a black hole mass of $10^9$ M$_{\bigodot}$ is $1.26 \times 10^{47}$ \lumcgs, we estimate the accretion rate to be $\sim 2-3$ per cent of the Eddington limit. The seed photon temperature of both \nthcomp\ components is found to be around 3--4 eV. This temperature is expected at a radius of $\sim 10$ \rg\ in a standard \cite{ss1973} accretion disc, given the black hole mass and accretion rate above. 
Finally, we show in Fig. \ref{fig:fluxes_wh} the flux of the hot \nthcomp\ component in the 3--10 keV band plotted against the flux of the warm \nthcomp\ component in the 0.3--2 keV band, for the different observations. Comparing these fluxes, we can probe the correlation between the primary emission (from the hot corona) and the soft excess. We find a Pearson's correlation coefficient of 0.72 with a $p$-value of 0.17.
Although this correlation is not highly significant, it suggests a trend of a stronger soft excess for an increasing primary flux.

We also add a caveat concerning the X-ray contribution from the jet. In principle, we cannot exclude a significant contribution from a non-thermal power law, which could be up to 70 per cent in the 2--10 keV band \cite[][]{grandi2007}. However, it is very unlikely that the X-ray emission is strongly contaminated by the jet. If that were the case, it would be difficult to explain the observed intensity of the \fek\ line \cite[]{wozniak1998,grandi2001}, the detection of a high-energy cut-off in the previous \sax\ and \nus\ observations, and the non-detection of gamma-rays. In particular, \cite{kataoka2011} reported an upper limit of \expo{5.2}{-12} \fluxcgs\ to the 0.1--10 GeV flux based
on two-year \fermi\ data and, from the broad-band spectral energy distribution, inferred that the jet contribution to the X-ray emission should be no more than a few per cent. 
\cite{hooper2016} later reported an even more stringent upper limit of \expo{2.2}{-12} \fluxcgs\ to the 0.1--100 GeV flux, from 85-month \fermi\ data.
This is consistent with a large jet angle to our line of sight \cite[]{giova2001}. However, a physical connection between the X-ray corona and the jet is certainly possible, as we discuss below.


\vspace{\baselineskip} 
 
\section{Discussion}\label{sec:discussion}
We reported on the broad-band high-energy (UV to hard X-rays) view of the radio/X-ray monitoring of the BLRG \tc\, based on five joint \xmm\ and \nus\ observations.

\tc\ appears to be weakly variable both in the UV and X-ray bands, at least on the time-scale of our campaign. Most of the flux variability is found in the soft X-ray band, whereas the flux and spectral variability at hard X-rays is small. Even if a significant correlation cannot be established, the data suggest a trend of a stronger soft X-ray flux corresponding to a larger UV flux. This is consistent with a Comptonization origin for the X-ray emission. The lack of spectral variability at hard X-rays (the photon index being always consistent with 1.78-1.79) would in turn suggest that the source is not undergoing significant geometrical or physical variations. Comparing with the two past \nus\ observations reported by \cite{ballantyne2014}, we find that the high-energy spectral shape is similar to Observation 1 (2012, higher flux) of \cite{ballantyne2014}, the photon indices being in good agreement with each other. The flux state, on the other hand, is intermediate between Observation 1 and Observation 2 (2013, lower flux) of \cite{ballantyne2014}, in which the 2--10 keV fluxes were found to be 5 and \expo{2.9}{-11} \fluxcgs, respectively.

We confirm the presence of a warm absorber, with physical parameters in rough agreement with the first detection reported by \cite{torresi2010}. However, from the analysis of the high-resolution RGS data, we find evidence for two distinct components: a high-ionization phase with $\log \xi \simeq 3$ and an outflow velocity of $\sim 2000$ \kms, and a low-ionization phase with $\log \xi \simeq 2$ and outflow velocity smaller than 600 \kms. 
We can derive an upper limit to the distance of such components \cite[see also][]{torresi2010}. By definition, $\xi = L_{\textrm{ion}}/ n R^2$, where $L_{\textrm{ion}}$ is the ionizing luminosity in the 1--1000 Ryd band, $n$ is the hydrogen gas density and $R$ the distance. Assuming that the gas is concentrated within a layer of thickness $\Delta r \leq R$, we have $R \leq L_{\textrm{ion}} / \nh \xi$, where $\nh = n \Delta r$. According to our best-fitting models, the average ionizing luminosity during the campaign is $\sim 1.5 \times 10^{45}$ \lumcgs. Then, we estimate $R \lesssim 100$ pc for the high-ionization component and $R \lesssim  1.5$ kpc for the low-ionization component. Both limits are consistent with that found by \cite{torresi2010} modelling the warm absorber as one-phase, i.e. $R \lesssim 60$ pc.
We can also derive a lower limit to the distance by assuming that the velocity $v$ of the outflow exceeds the escape velocity, implying that $R \geq 2 G M_{\textrm{BH}}/v^2$ where $G$ is the gravitational constant. We estimate $R \gtrsim 1.5$ pc for the high-ionization (and high-velocity) component and $R \gtrsim 100$ pc for the low-ionization component. The warm absorber is thus consistent with originating from beyond the broad-line region and the torus, which are likely located within 0.06 pc and 1.5 pc respectively \cite[]{torresi2010}. The high-ionization/high-velocity component is likely launched closer to the AGN. 

We also confirm the presence of a neutral \fek\ emission line not accompanied by a significant Compton reflection component.
The line is consistent with being constant during our campaign, and is also consistent with being narrow in at least three observations out of five. However, the line is apparently broad in observation 3 and (marginally) in observation 4. The interpretation of these results is not straightforward. If the line broadening is interpreted as a signature of relativistic effects from the accretion disc, it would imply an origin of the line within 20 \rg\ of the black hole during observation 3 only. The lack of a Compton hump, however, would be difficult to explain in this case. Indeed, a self-consistent modelling with \relxill\ indicates that the putative reflecting material would require quite extreme parameters, such as an iron overabundance larger than 7 times to solar and a small covering fraction. We note that \cite{sambruna2011} detected five emission lines in the Fe K band of the 2007 \suz\ spectrum (with pretty much the same flux level as our campaign), and found that two ionized reflection components are needed to reproduce such features. Puzzling enough, the neutral \fek\ line was found to have a width of \aer{0.118}{0.020}{0.013} keV when modelled with a Gaussian \cite[]{sambruna2011}; this value is consistent with our observation 4 (0.13 keV) but not with observation 3 ($>0.2$ keV). 
On the other hand, the \fek\ line is narrow (and relatively weak) in the 2008 \xmm/pn spectrum, consistently with past observations with previous instruments \cite[]{wozniak1998,grandi2001}, albeit with some ambiguity related to the spectral modelling \cite[]{wozniak1998}. We cannot thus exclude that the line broadening in observation 3 is due to instrumental effects. 
However, according to our results, the \fek\ line could originate from a large-scale structure, such as the obscuring torus or the broad-line region, with a column density of around \expo{6}{23} \sqcm. 
Finally, the lack of a disc reflection component could also indicate that the corona is outflowing rather than being static, so that the X-ray emission could be beamed away from the disc \cite[e.g.][]{belo1999,malzac2001}. As noted by \cite{ballantyne2014}, according to the model of \cite{malzac2001} a corona outflowing with a velocity of $\sim 0.5 c$ would be consistent with both the observed photon index and the small reflection fraction. Given the presence of a relativistic radio jet in \tc, it is conceivable that the X-ray corona could form the base of the jet itself. The corona/jet connection will be investigated in a forthcoming paper, in which we will present the radio data of the campaign.

In agreement with previous observations, we find the presence of a significant soft X-ray excess below 2 keV. 
Ionized reflection is not able to self-consistently explain the soft excess, the \fek\ line and the lack of a Compton hump in this source. On the other hand, a warm Comptonization model is found to well describe the soft excess. Its photon index is found to be around 2.4-2.5, while the temperature is consistent with 0.6 keV, corresponding to an optical depth of around 20. Interestingly, these parameters are in agreement with those generally found in radio-quiet Seyferts, although most radio-quiet sources tend to show lower temperatures between 0.1 and 0.4 keV \cite[]{cheeses,jin2012a}. From the best-fitting parameters, we can compute the Compton amplification factor $A_w$, namely the ratio between the total power emitted by the warm corona and the seed soft luminosity from the accretion disc. Following the procedure described in \cite{cheeses}, we estimate an amplification $A_w \simeq 2$, i.e. the value theoretically expected for a slab corona fully covering a passive disc \cite[see also][]{pop2013mrk509}. The soft excess is thus consistent with originating in a slab-like, warm and optically thick corona which could form the upper layer of a nearly passive accretion disc \cite[see also][]{rozanska2015}. This in turn suggests an intriguing similarity between \tc\ and radio-quiet Seyferts, and supports the idea that the same mechanism regulates accretion in luminous radio-loud (jetted) AGN and in average radio-quiet (non-jetted) AGNs.

Concerning the hot corona, we derive only lower limits to the temperature of 20--40 keV. This corresponds to an upper limit to the optical depth of around 4.
We also estimate the amplification factor $A_h$ of the hot corona to be around 20 \cite[only weakly depending on the temperature; see][]{cheeses}. This large value indicates a compact or patchy corona, intercepting only a fraction of the seed soft photons. Indeed, the geometrical parameter $g$ describing the compactness or ``patchiness'' of the corona can be estimated as $\sim 2/A_h$ \cite[]{cheeses}. We thus find $g\simeq 0.1$, in excellent agreement with the result independently obtained by \cite{ballantyne2014}. Our results are thus consistent with a ``two-corona'' scenario in which the UV/soft X-ray emission is produced by an extended warm corona and the hard X-ray emission is due to a more compact, hot corona that intercepts roughly 10\% of the seed soft photons. 
Some ambiguity remains concerning the precise geometry of the hot corona, which could be patchy and possibly outflowing, or fill the inner part of the accretion flow in a truncated disc scenario. In the latter case, the ``cold phase'' of the accretion flow (disc/warm corona) would extend down to a transition radius, while the hot corona would fill the space down to the innermost stable circular orbit around the black hole. With this geometry, the reflection features are expected to be weak, since the cold phase subtends a small solid angle from the hot corona. Such a geometry has been widely applied to X-ray binaries in the hard/jetted state \cite[e.g.][and references therein]{malzac2016}, but also to radio-quiet Seyferts \cite[for a detailed discussion, see][]{pop2013mrk509}.
Finally, our results suggest that the soft excess could be related to the primary X-ray emission (Fig. \ref{fig:fluxes_wh}), in turn indicating a physical relationship between the warm and hot coronae. This is clearly expected if the warm corona is indeed the upper layer of the disc, as the seed photons entering the hot corona would actually originate in the warm corona.

The results from our campaign on \tc\ can be compared with those from an analogous \xmm/\nus\ monitoring program on the radio-quiet Seyfert 1 \ngc\ ($5\times 20$ ks simultaneous observations spaced by 2 days), carried out in 2015 \cite[]{4593}. This has been the first \xmm/\nus\ campaign explicitly aimed at studying the high-energy emission of an AGN by probing its variability on time-scales of days. \ngc\ is an X-ray bright Seyfert 1 galaxy hosting a supermassive black hole of $\sim 10^7$ solar masses \cite[][]{mbh4593}, i.e. two orders of magnitude smaller than \tc.
The campaign on \ngc\ has shown the source to be remarkably variable, both in flux and spectral shape, on time-scales as short as a few ks, with a softer-when-brighter behaviour. From the X-ray spectral analysis, the primary power law was found to have a photon index varying between 1.6 and 1.8 and a high-energy cut-off varying from \aer{90}{20}{40} keV to \linf{700} keV. Moreover, the high-energy data are consistent with the presence of two distinct reflection components producing a significant \fek\ line with a broad component, and a moderate Compton hump. \tc, on the other hand, is much less spectrally complex and variable.
However, the properties of the soft excess are remarkably similar between the two objects. First, in both sources the soft excess is not explained by ionized reflection alone, thus favouring a warm Comptonization scenario. Moreover, the soft excess is correlated with the primary emission in \ngc, and we observe a similar trend in \tc. These results suggests that the two-corona scenario could provide a viable physical model for both sources, in spite their belonging to different classes of AGNs. The observational differences (most notably the presence/lack of a jet) could be due to different geometrical and/or physical parameters, such as a different accretion rate and the presence of large-scale magnetic fields (likely needed to accelerate and collimate jets).
Further observational and theoretical work will be needed to fully explore the properties and implications of the two-corona scenario for the high-energy emission of AGNs.

\section*{Acknowledgements}
We are grateful to the referee for helpful comments that improved
the manuscript.
This work is based on observations obtained with: the \nus\ mission, a project led by the California Institute of Technology, managed by the Jet Propulsion Laboratory and funded by NASA; \xmm, an ESA science mission with instruments and contributions directly funded by ESA Member States and the USA (NASA).
This research has made use of data, software and/or web tools obtained from NASA's High Energy Astrophysics Science Archive Research Center (HEASARC), a service of Goddard Space Flight Center and the Smithsonian Astrophysical Observatory. 
The research leading to these results has received funding from the
European Union's Horizon 2020 Programme under the AHEAD project (grant
agreement n. 654215).
FU acknowledges financial support from the Italian Space Agency (ASI) under contract ASI/INAF 2013-025-R01.
SB and MC acknowledge financial support from ASI under grant ASI-INAF I/037/12/0, while GM, AM and AT from ASI/INAF I/037/12/0-011/13. 
SB, GM, AM, RM and AT acknowledge financial support from the European Union Seventh Framework Programme (FP7/2007-2013) under grant agreement no. 312789.
FU, GM, SB, MC, MD, PG, ET, ADR, MG, AM, RM and AT acknowledge financial contribution from the agreement ASI-INAF n. 2017-14-H.O.
BDM acknowledges support from the Polish National Science Center grant Polonez 2016/21/P/ST9/04025.
GP ackowledges support by the Bundesministerium f\"ur Wirtschaft und Technologie/Deutsches Zentrum f\"ur Luft und Raumfahrt (BMWI/DLR, FKZ 50 OR 1408) and the
Max Planck Society.
\bibliographystyle{mnras}
\bibliography{mybib.bib}

\end{document}